\documentclass[12pt]{iopart}
\usepackage{graphicx}
\usepackage{float}
\usepackage{physics}
\usepackage[pdftex]{color}
\usepackage{hyperref}
\usepackage{xcolor}
\definecolor{light-gray}{gray}{0.95}
\newcommand{\code}[1]{{\texttt{#1}}}
\usepackage{amsmath}

\usepackage[
backend=biber,
style=ieee,
]{biblatex}

\begin{document}

\def\hilite#1{\textbf{\color{red}{#1}}}

\title{Design of Passive and Structural Conductors for Tokamaks Using Thin-Wall Eddy Current Modeling}

\author{A.F. Battey$^1$, C. Hansen$^1$, D. Garnier$^2$, D. Weisberg$^3$, C. Paz-Soldan$^1$, R. Sweeney$^4$, R.A. Tinguely$^2$, A.J. Creely$^4$}

\address{$^1$ Department of Applied Physics and Math, Columbia University, New York 10027,
United States of America}
\address{$^2$ Massachusetts Institute of Technology, Cambridge, Massachusetts 02139, United States of America}
\address{$^3$ General Atomics, San Diego, California 92121, United States of America}
\address{$^4$ Commonwealth Fusion Systems, Devens, Massachusetts 01434, United States of America}

\ead{a.battey@columbia.edu}
\vspace{10pt}
\begin{indented}
\item[January 2023]
\end{indented}

\begin{abstract}
A new three-dimensional electromagnetic modeling tool (\code{ThinCurr}) has been developed using the existing PSI-Tet finite-element code in support of conducting structure design work for both the SPARC and DIII-D tokamaks. Within this framework a 3D conducting structure model was created for both the SPARC and DIII-D tokamaks in the thin-wall limit. This model includes accurate details of the vacuum vessel and other conducting structural elements with realistic material resistivities. This model was leveraged to support the design of a passive runaway electron mitigation coil (REMC), studying the effect of various design parameters, including coil resistivity, current quench duration, and plasma vertical position, on the effectiveness of the coil. The REMC is a non-axisymmetric coil designed to passively drive large non-axisymmetric fields during the plasma disruption thereby destroying flux surfaces and deconfining RE seed populations. These studies indicate that current designs should apply substantial 3D fields at the plasma surface during future plasma current disruptions as well as highlight the importance of having the REMC conductors away from the machine midplane in order to ensure they are robust to off-normal disruption scenarios.

\end{abstract}


\section{Introduction}
\label{sec:Intro}

Large currents can be induced in tokamak conducting structures in the presence of time-varying magnetic fields, which can be produced by the plasma or external coils. For example, significant induced magnetic fields are generated in the majority of tokamak discharges during the plasma start-up and breakdown phase when currents are ramped quickly in the equilibrium field coils \cite{maxwell_james_c_faradays_1855}. Changing magnetic fields are also present in tokamak disruptions during the current quench (CQ) as the plasma current decays rapidly due to a sudden increase in the plasma resistivity after the thermal quench phase\cite{wesson_disruptions_1989}. These changing magnetic fields have the ability to generate large electric fields which when coupled with non-axisymmetric features in nearby conducting structures have the potential to drive large non-asymmetric currents and therefore produce large non-axisymmetric magnetic fields. Non-axisymmetric magnetic fields can have a variety of negative consequences from affecting the null in the equilibrium poloidal field\cite{pustovitov_models_2022}, needed for plasma breakdown, to inducing large forces on structures, and potentially driving dangerous magnetohydrodynamic (MHD) instabilities\cite{jackson_iter_2008, jackson_simulating_2009, leuer_plasma_2010}. However, if properly accounted for, these three-dimensional currents can also be beneficial by slowing the growth rates of MHD instabilities \cite{boozer_simplified_2010,clement_gpu-based_2017, clement_optimal_2018, mauel_dynamics_2005, okabayashi_control_2005, battey_simultaneous_2023} and deconfining runaway electrons (REs) in the case of runaway electron mitigation coils (REMCs), which are being designed for both the SPARC \cite{izzo_runaway_2022, tinguely_minimum_2023, tinguely_modeling_2021, sweeney_mhd_2020, creely_overview_2020, rodriguez-fernandez_overview_2022, izzo_analysis_2012, greenwald_status_2020} and DIII-D \cite{luxon_design_2002, commaux_novel_2011, hollmann_experiments_2010, eidietis_control_2012, taylor_disruption_1999, shiraki_dissipation_2018, weisberg_passive_2021} tokamaks. It will  be demonstrated in this work that these  eddy currents can also significantly affect the response time  and efficacy of these REMCs. These potential negative consequences are expected to become more significant in future fusion-relevant machines which will have larger equilibrium magnetic fields and plasma currents, and will therefore be more prone to runaway electrons (REs) and large vessel forces. Therefore, in order to accurately and reliably design and predict the behavior of future devices, it is crucial to be able to simulate the three-dimensional behavior of currents induced in device conducting structures \cite{iter_physics_expert_group_on_disruptions_chapter_1999, lehnen_disruptions_2015, breizman_physics_2019, boozer_two_2011, boozer_runaway_2017}.

\par
The remainder of this paper will be split into five sections. In Section~\ref{sec:PsiTet}, the physics and numerical methods employed by \code{ThinCurr} are discussed. A study of the proposed REMCs for both the DIII-D and SPARC devices is then presented in Section~\ref{sec:REMC} before a discussion of disruption-induced forces in Section~\ref{sec:Forces}. Conclusions and ideas for future work are the discussed in Section~\ref{sec:Conclusion}.

\section{ThinCurr: A New 3D Thin-Wall Eddy Current Modeling Tool}

\label{sec:PsiTet}

\begin{figure}[H]
\begin{center}
    \includegraphics[height=7cm]{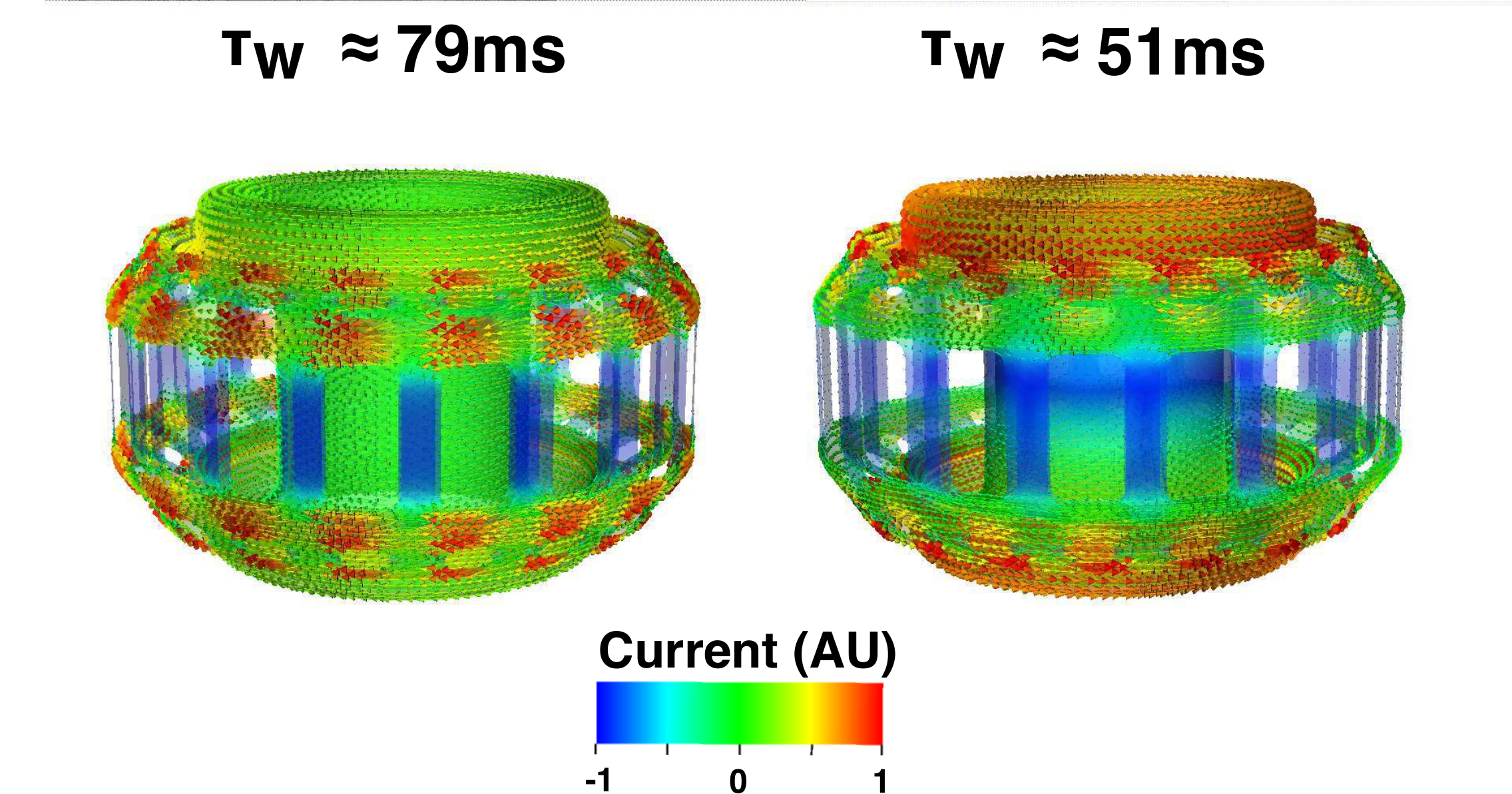} 
    \caption{The structure and current decay time-scales of the longest lived (left) and top-down anti-symmetric (right) current patterns for the SPARC vacuum vessel as calculated by the \code{ThinCurr} code.}
    \label{eigenVal_SPARC}
\end{center}
\end{figure}

This work utilizes a newly-developed 3D electromagnetic modeling code (\code{ThinCurr}), that is based on the PSI-Tet 3D MHD code~\cite{hansen_mhd_2014, hansen_numerical_2015, boozer_equations_1998}, and solves a thin-wall eddy current formulation, similar to that used in the \code{VALEN}~\cite{bialek_modeling_2001} and \code{STARWALL}~\cite{merkel_linear_2015} codes, that has proven useful for a variety of electromagnetic applications in fusion \cite{okabayashi_control_2005, mauel_dynamics_2005,katsuro-hopkins_enhanced_2007, hanson_validation_2016, clement_gpu-based_2017, clement_optimal_2018, boozer_simplified_2010, bialek_modeling_2001, boozer_robust_2004, boozer_resistive_2003, boozer_stabilization_1995, battey_simultaneous_2023, boozer_equations_1998}. However, \code{ThinCurr} leverages the modern features of PSI-Tet, such as a CAD interface, the use of unstructured (triangular) grids, parallelization, and interfaces to scalable direct and iterative solvers, to provide additional power and flexibility over prior tools used to study eddy currents in fusion devices. This section provides a brief overview of the code and methods, sufficient to support the presented results. A separate publication will follow with additional information on details of numerical methods and features beyond the scope of this work.

\par
Due to its flexible unstructured finite-element representation, it is straightforward to develop detailed models for 3D conducting structures based on high fidelity CAD models of existing and future machines. The work presented in this article will highlight analysis work completed for both the DIII-D \cite{luxon_design_2002} and SPARC tokamaks \cite{creely_overview_2020, greenwald_status_2020}. An example of the effect of 3D features on calculated current decay time-scales is highlighted in Figure \ref{eigenVal_SPARC} which shows the current distribution (eigenvector) corresponding to the largest eigenvalue ($L\dot{I} = \tau_W RI$) for the SPARC vacuum vessel model computed by \code{ThinCurr}. The eigenvector pictured on the left corresponds to the longest lived current pattern, which for a simple torus corresponds to an axisymmetric toroidal current localized on the outboard midplane.  Without accounting for three-dimensional features, this current decay time-scale is significantly overestimated by $49.4\%$  when compared to the more accurate 3D model (118ms (2D) vs 79ms (3D)) shown in figure \ref{eigenVal_SPARC}~(left). The shorter time-scale in the 3D case is a consequence of currents being forced to take longer and narrower, and therefore more resistive, current paths as they maneuver around 3D features such as vessel ports.

\par
This difference between the 2D approximation and the full 3D model can be less significant for particular eigenvectors such as the top-bottom anti-symmetric mode shown in Figure \ref{eigenVal_SPARC}~(right). This difference (60ms (2D) vs 51ms (3D)) is less than the largest eigenvalue case as the majority of the current flows in the top and bottom of the vessel. These current paths are therefore not forced to change as significantly in the presences of vessel ports, which are primary located at the midplane, and therefore experience comparable resistivities to the 2D approximation. This particular current pattern is important as its decay affects the growth rate of the vertical instability and limits the time-response of external coils in toroidal devices.

\par
For the DIII-D REMC, \code{ThinCurr} was used to determine the effect of coil radius, and therefore distance from the center-post, on the coil's performance. The effect of coil resistivity was then explored for the REMCs on both devices. The design robustness of both coils was also verified for the case a varying CQ duration and plasma vertical position during the CQ which could vary during a `hot' vertical displacement event (VDE).

\subsection{Model Description}

\code{ThinCurr} computes the dynamics of currents, subject to inductive coupling and resistive dissipation, on one or more arbitrarily shaped 3D surfaces, corresponding to a thin-wall approximation of conducting regions, and closed filaments. On each surface the current is represented as the rotated gradient of a scalar potential function,
\begin{equation}
\textbf{J}_s = \textbf{J} * t_w = \nabla \chi \times \hat{n},
\end{equation}
where $t_w$ and $\hat{n}$ are the thickness and unit normal to the surface, respectively. The scalar potential $\chi$ is discretized using Lagrange finite elements (FEs) on a triangular mesh. \code{ThinCurr} supports high order finite element representations, but for this work only linear elements were used. Additional elements are also added to the single-valued Lagrange FE space to enable representation of a multi-valued $\chi$ as needed by surfaces with holes and other multiply connected topology. 

An L-R circuit model is then defined,
\begin{equation}\label{eq:thincurr_lr}
L_{ij} \dot{I}_j + R_{ij} I_j = V_i(t),
\end{equation}

where each ``circuit" ($i$) corresponds to a single basis function in the spatial discretization, whose current forms a divergence free loop by construction. The inductance and resistance matrices can then be computed as
\begin{equation}\label{eq:thincurr_L}
L_{ij} = \int \int \frac{\left(\nabla \chi_i \times \hat{n}\right) \cdot \left(\nabla \chi_j \times \hat{n}\right)}{\left| r' - r \right|^2}  dA' dA
\end{equation}
and
\begin{equation}
R_{ij} = \int \eta \left(\nabla \chi_i \times \hat{n}\right) \cdot \left(\nabla \chi_j \times \hat{n}\right) dA
\end{equation}
respectively, where the resistivity $\eta$ is defined as a piecewise constant function. Resistivity is allowed to vary between regions in the mesh, which consist of any arbitrary non-overlapping groupings of triangles that are defined via the mesh pre-processor (eg. CUBIT). For the linear elements used in this work, an analytic solution~\cite{ferguson_complete_1994} for the potential ($\phi(r) = \int \frac{1}{\left| r - r' \right|^2} dA'$) from a triangular element is used to treat the singularity in equation~\ref{eq:thincurr_L}. 

For currents defined as filaments, the Biot-Savart law is used to replace the inner integration in equation~\ref{eq:thincurr_L}. A user-defined radius and resistivity per unit length are used separately to determine self-inductance and resistivity of the filament respectively. Filaments can be defined by a major radius and vertical position pair, for circular coils, or by a collection of 3D points in Cartesian space for arbitrarily-shaped coils. Multiple filaments can be grouped together to represent series wound coils where the current through all windings must be the same. Voltages or currents are then specified in these filaments, which form the source term on the RHS of equation~\ref{eq:thincurr_lr}.

Several solution types for equation~\ref{eq:thincurr_lr} are of interest for conducting structures, primarily the vacuum vessel, in tokamaks and other fusion reactors, including: 1) Time-domain simulations with given initial conditions and $V(t)$ (e.g. disruption simulations), 2) Frequency-domain simulations for the steady state response to oscillatory $V(t) \propto e^{i\omega t}$ (e.g. predicted sensor signals from rotating MHD modes), and 3) Eigenvalue calculations for the characteristic timescales ($\tau_{L/R}$) and modes of the conducting structures (e.g. vertical stability). Both direct and iterative solvers are provided for each type of solution, with parallelized versions available either through libraries (eg. MKL) or internally through PSI-Tet's OpenMP+MPI parallelization. For time-domain calculations a direct solver is often the most efficient approach as the factorization cost is amortized over many time steps, while eigenvalue and frequency-response calculations usually benefit from iterative methods as often only a handful of eigenvalues or a single solve at a given frequency are desired.

For all problem types additional sensors can be defined to measure the magnetic flux through arbitrary closed loops and current flowing across curves in the mesh (cross-sections in the thin-wall limit). Additionally, plot files are saved using HDF5 \cite{noauthor_hdf5_nodate} and Xdmf \cite{noauthor_xdmf_nodate} formats to enable use of VisIt \cite{noauthor_citing_nodate} and other tools, to help visualize the 3D structure of currents, magnetic fields, and forces.

\par
\begin{figure}[H]
\begin{center}
    \includegraphics[height=9cm]{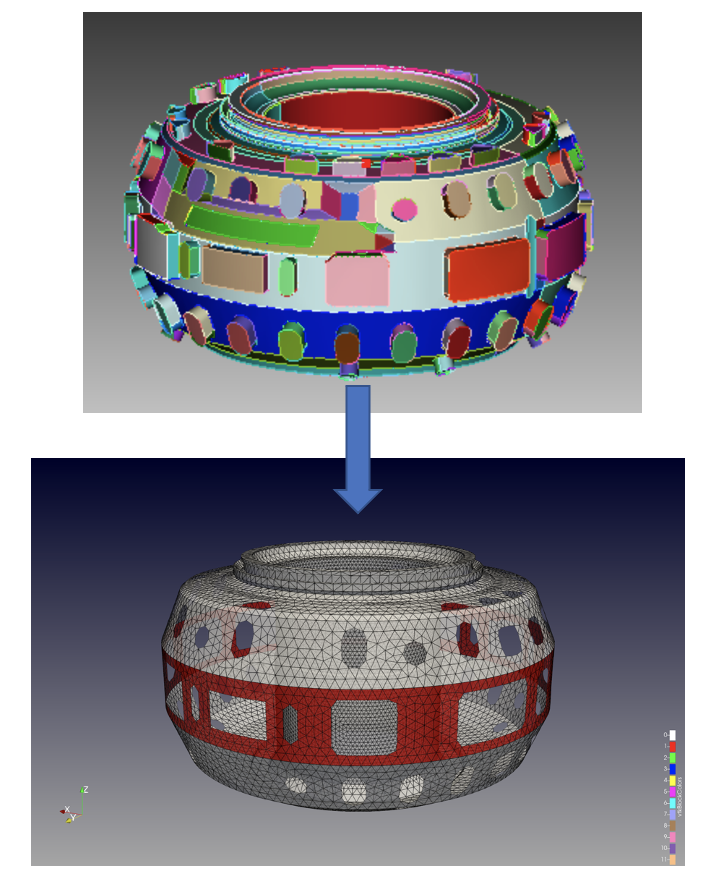} 
    \caption{Example of complexity of the original DIII-D CAD model (top) and the \code{ThinCurr} DIII-D modeling (bottom) following defeaturing to only retain the features which are most important to the phenomenon of interest. The colors in the lower plot represent `blocks' which allow materials with different resistivity or material thickness to be defined.}
    \label{defeaturing}
\end{center}
\end{figure}

\code{ThinCurr} can be coupled with professional CAD software to precisely capture important details of potentially complex 3D conductors. For the case of the DIII-D and SPARC tokamaks, their engineering teams maintain highly detailed and accurate CAD models which contain features too small to be captured by a reasonable mesh resolution. Therefore, the first step of developing a \code{ThinCurr} model is to determine which features will be the most significant for the phenomenon of interest. For the study of disruption and start-up vessel currents this often includes large vessel ports of 3D structures capable of carry significant currents. The remaining features are then removed with a comparison between the full DIII-D CAD model and the defeatured \code{ThinCurr} model shown in Figure \ref{defeaturing}. 

\par
\begin{figure}[H]
\begin{center}
    \includegraphics[height=5cm]{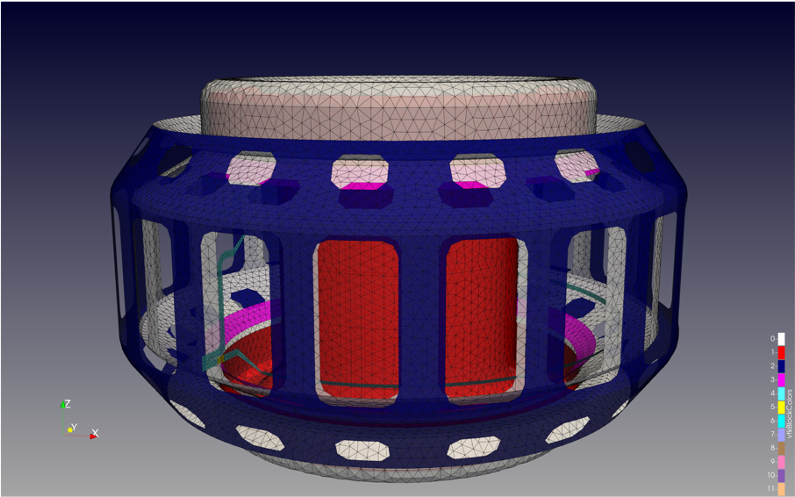} 
    \caption{A visualization of the SPARC \code{ThinCurr} model where each color represents a different modeled resistivity region. These different resistivities are a result of either different thicknesses or materials. }
    \label{SparcMaterials}
\end{center}
\end{figure}

Regions and individual conductors can then be separately defined to allow different material resistivities to be defined. These differences allow for different materials to be modeled as well as allow the thickness of the conductors to vary across the model. This is the case for both the DIII-D and SPARC vacuum vessel models which are uniform in material but have thicknesses that vary from the high-field to low-field side. An example of these defined regions can be seen in Figure \ref{defeaturing} for DIII-D, which has a vessel which is significantly thicker on the low-field side midplane. The thickness of the SPARC vacuum vessel also varies significantly as pictured in Figure \ref{SparcMaterials} as well as the SPARC REMC which is made of a different material. 

\par
The model is then triangularly meshed to the desired resolution where it is possible to mesh different regions to higher or lower resolution. This ability was implemented for the DIII-D model whose center-post is modeled with significantly higher resolution due to its proximity to the REMC. Following the meshing all holes in the model must be defined to ensure all flux is accounted for and the final set of equations is well-posed. For example, a simple closed torus would required two holes to be defined: one for the toroidal flux and one for the poloidal flux. These holes correspond to loops around the minor and major circumference, respectively. Trapped volumes such as those created by a sphere or torus also require a `closure mesh' which corresponds to the removal of an equation from the final equation set \cite{bialek_modeling_2001}. For the case of a closed torus only a single closure is required.

\section{Runaway Electron Mitigation Coil Design}
\label{sec:REMC}

\par
\code{ThinCurr} was also leveraged to explore the effectiveness of proposed runaway electron mitigation coils (REMCs) for both the DIII-D and SPARC tokamaks. These coils are passive structures which are designed to prevent and mitigate runaway electrons (REs) which are produced by large loop voltages driven during the CQ phase of tokamak disruptions \cite{boozer_two_2011, smith_passive_2013}. These large electric fields are capable of producing REs through two primary mechanisms \cite{breizman_physics_2019}. Electrons flowing through a hot plasma experience a collisional drag force which is a decreasing function of velocity ($1/v_e^2$) for electrons moving faster than the electron thermal velocity and this force asymptotes for ultra-relativistic electrons at: 

\begin{equation}
\frac{4\pi e^4 n \ln{\Lambda}}{mc^2},
\end{equation}
where $n$ is the electron density, $m$ is the electron mass, $e$ is the electron charge, $c$ is the speed of light, and $\ln{\Lambda}$ is the Coulomb logarithm. Therefore, application of an electron field in excess of the Connor-Hastie field,

\begin{equation}
E_{CH} = \frac{4\pi e^3 n \ln{\Lambda}}{mc^2},
\end{equation}
will eventually produce runaway electrons with unimpeded acceleration of electrons with $v>c\sqrt{E_c/E}$ \cite{connor_relativistic_1975, dreicer_electron_1959, dreicer_electron_1960}. Only a small fraction of the entire electron population falls immediately into the runaway energy range when the electric field is less than the Dreicer field,

\begin{equation}
E_{D} = \frac{4 \pi e^3 n \ln{\Lambda}}{T},
\end{equation}
unless there is a significant hot tail on the electron distribution function.\cite{svenningsson_hot-tail_2021, smith_hot_2008,aleynikov_generation_2017, hollmann_study_2017} This hot tail is commonly formed in tokamak disruption plasmas due to an influx of impurities which causes cooling of the hot electrons through collisional drag. Since the higher energy electrons experience fewer collisions they are cooled at a slower rate resulting in surviving ``hot-tail" of the initial electron Maxwellian which is strongly susceptible to runaway. The combination of this hot-tail effect and Dreicer production are the primary mechanism for runaway production with these effects dominating in current tokamaks. 

\par
The secondary or avalanche mechanism occurs when a thermal electron experiences a hard collision with an existing runaway electron and is bumped into the runaway population while the initial runaway remains in the relativistic population. This effect results in a runaway population which grows exponentially with time as opposed to a linear growth induced by the Dreicer effect \cite{rosenbluth_theory_1997}. The efficiency of this mechanism scales as an exponential function of the predisruption plasma current and will therefore dominate along with the hot-tail mechanism for future devices like ITER and SPARC \cite{sweeney_mhd_2020, boozer_runaway_2017, saint-laurent_overview_2013, nilsson_kinetic_2015,mcdevitt_avalanche_2019, martin-solis_formation_2017, hesslow_influence_2019}. In fact this mechanism is expected to be so efficient that it is expected that a considerable fraction of plasma current will be converted directly to RE current unless avoidance methods are implemented.

\par
These RE beams will potentially carry MAs of current, composed of electrons in the MeV range of energies, in future large plasma current devices such as ITER and SPARC. This will pose a serious risk to plasma facing components in future devices. At high plasma currents in ITER RE impacts on the Be limiters could lead to water leaks \cite{m_lehnen_rd_2018, reux_runaway_2015}. SPARC is designed for shorter 10s plasma flattops and does not have active cooling of the tungsten first wall tiles, limiting the worst RE impact failure mode to significant melting \cite{creely_overview_2020}. Work remains to assess the space of possible melt damage.  Proposed REMCs must be designed to couple to the large disruption-induced loop voltages and apply significant non-axisymmetric fields capable of safely deconfining REs and preventing further RE seed generation. Screening from induced eddy currents, as well as the forces on the overall system, are a crucial factors to consider when evaluating the performance of a potential REMC. 

\par
A large effort is therefore underway to determine a method of avoiding REs and, if avoidance is not possible, a method by which the deleterious effects of REs can be safely mitigated \cite{hender_chapter_2007, lehnen_disruptions_2015, hollmann_status_2014, boozer_theory_2015}. An exciting candidate for both RE avoidance and mitigation is a passive REMC which takes advantage of the large loop voltage generated by the disruption CQ by placing a non-axisymmetric conductor inside the tokamak vacuum vessel. If this coil is designed properly this loop voltage will drive large currents in the passive conductor which will in turn produce significant non-axisymmetric magnetic fields. Similar 3D fields have been shown both numerically and experimentally to deconfine relativistic electrons by destroying magnetic surfaces \cite{papp_effect_2013, papp_runaway_2011, gobbin_runaway_2017, sommariva_electron_2018}, driving large MHD instabilities \cite{zeng_experimental_2013, izzo_analysis_2012, aleynikov_theory_2015, paz-soldan_kink_2019, paz-soldan_novel_2021, liu_mars-f_2019, reux_demonstration_2021}, and broadening the pitch angle distributions \cite{lehnen_disruptions_2015, hollmann_status_2014, chen_suppression_2018, boozer_two_2011, smith_passive_2013, jiang_simulation_2016, mlynar_runaway_2018, papp_effect_2013}. These mechanisms are capable of both preventing the formation of a RE-beam by deconfining seed populations and destroying magnetic surfaces as well as terminating an existing population. Therefore, the effectiveness of the coil is dependent on applying a large 3D field early in the CQ. As a consequence of this, it is crucial to understand potential screening current induced in conductors near the coil which can significantly reduce the magnitude of the applied field at this critical stage. 

\begin{figure}[H]
\begin{center}
    \includegraphics[height=5cm]{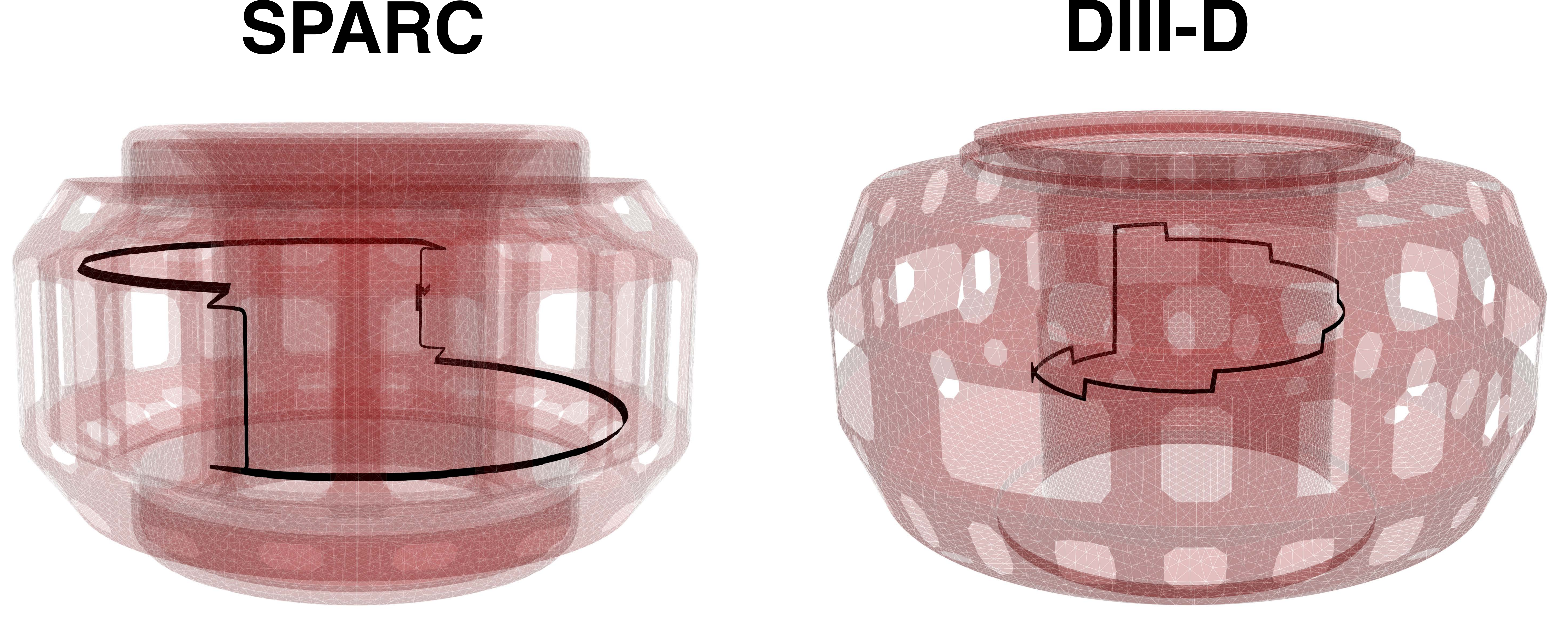} 
    \caption{A visualization of the SPARC (left) and DIII-D (right) \code{ThinCurr} vacuum vessel model in red with the proposed REMC models visualized in black.}
    \label{REMCcompare}
\end{center}
\end{figure}
\par
As mentioned previously, plasma disruptions are not the only time where large loop voltages are present during a tokamak plasma discharge such as those applied during plasma start-up or from the vertical stability coils (VSCs), which are expected to be located very near the SPARC REMC. There is therefore a risk of the proposed REMC reacting to these various loop voltages and apply unwanted 3D fields. For this reason the REMCs will be able to be placed into an `open-circuit' mode which will prevent current from flowing in the coils. When desired the coils can be returned to their nominal states, where current does not flow until a threshold voltage is reached. This allows passive activation of the coil, while ensuring non-axisymmetric fields are only applied when desired. The REMCs will also be equipped with a variable resistor which will allow both the time-response and maximum voltage to be tailored for a specific research goal.

\begin{figure}[H]
\begin{center}
    \includegraphics[height=7cm]{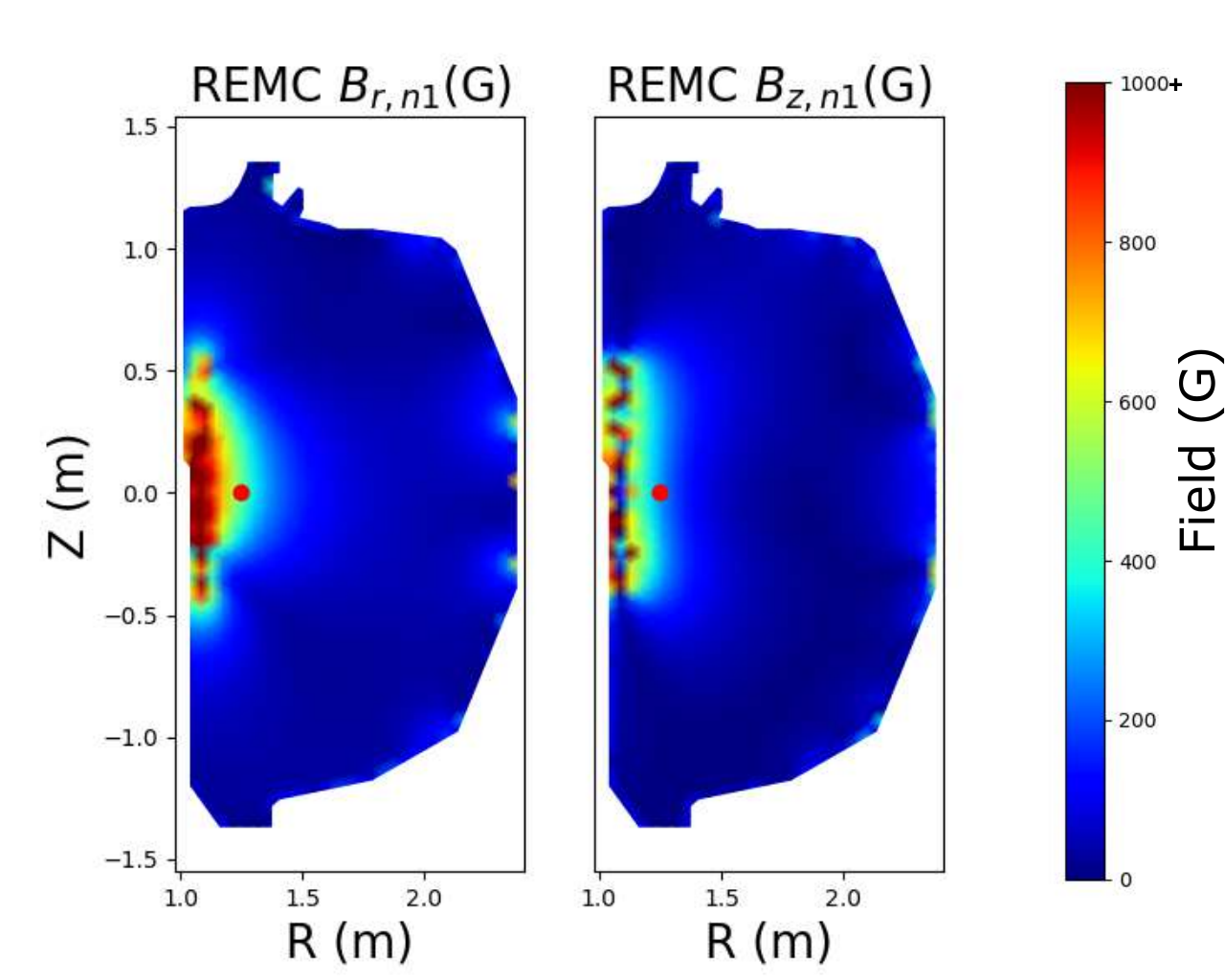} 
    \caption{The $n=1$ field applied by the DIII-D REMC at $t=18ms$ during a full plasma current disruption with $\tau_{CQ}= 12ms$, where is the CQ begins at $t=1ms$. The red dot shows the location of the magnetic field measurements shown in later figures.}
    \label{DIIID_fields}
\end{center}
\end{figure}

\begin{figure}[H]
\begin{center}
    \includegraphics[height=7cm]{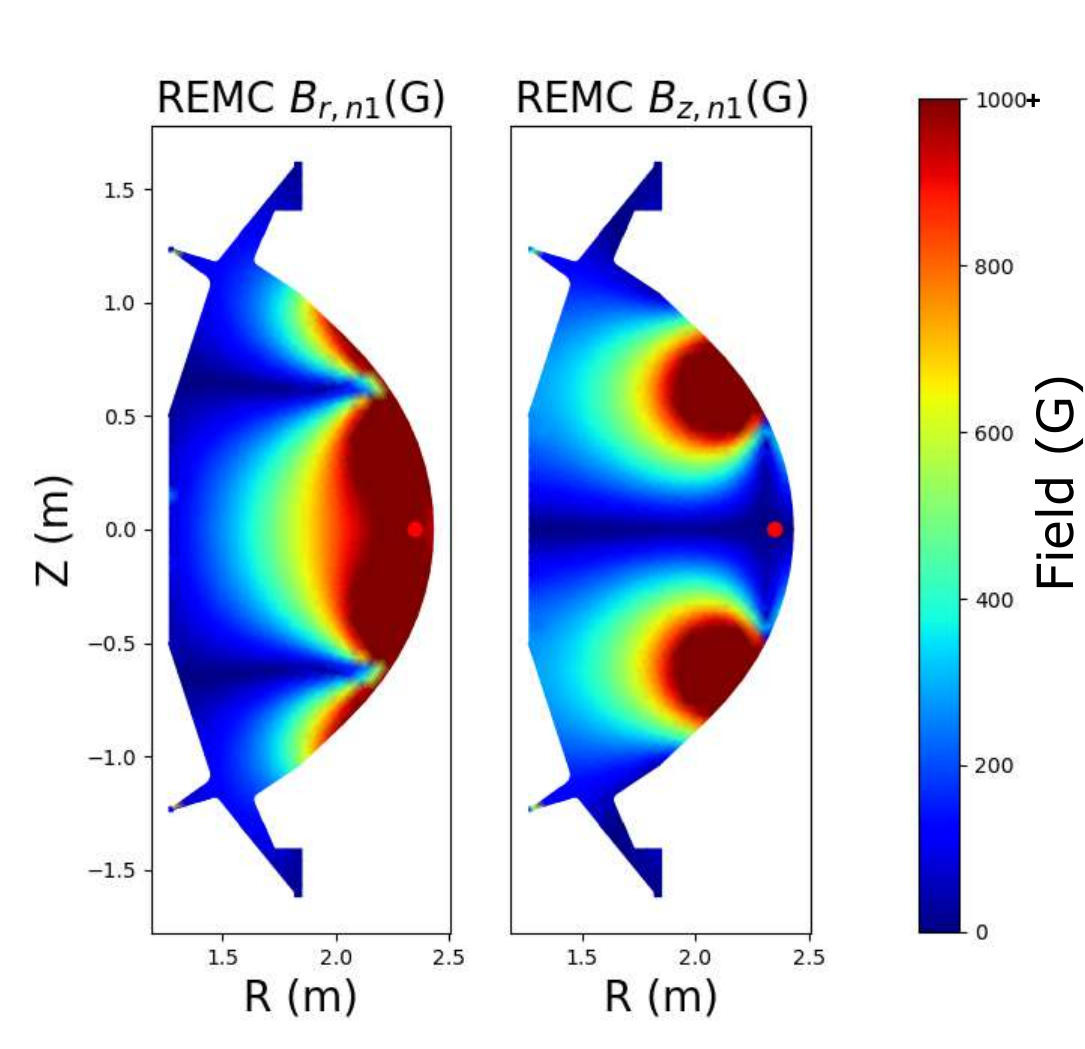} 
    \caption{The $n=1$ field applied by the SPARC REMC at $t=5ms$ during a full plasma current disruption with $\tau_{CQ}= 3.2ms$, where is the CQ begins at $t=1ms$. The red dot shows the location of the magnetic field measurements shown in later figures.}
    \label{SPARC_fields}
\end{center}
\end{figure}

\par
The \code{ThinCurr} code was leveraged to study the effect of screening currents induced in conducting structures near both the DIII-D and SPARC REMCs as well as the ability to inductively couple to the changing plasma current. These screening current effects can vary significantly between devices due to substantial differences in the vacuum vessel resistivities. While both coils are similar in that they are designed to drive a primarily n=1 field structure, they differ significantly because the DIII-D coil will be mounted on the high-field (inboard) side near the midplane while the SPARC coil will be attached to the low-field (outboard) side near the Vertical Stability Coils (VSCs) as shown in Figure~\ref{REMCcompare}. These different coil shapes also result in significantly different applied field structures with the radial and vertical fields for the DIII-D and SPARC REMCs shown in Figure \ref{DIIID_fields} and \ref{SPARC_fields}, respectively. As a result of these design choices this paper will also serve as a comparison between these two potential mounting locations. Since the performance of the coil depends crucially on its ability to couple to changing currents within the plasma, special attention was paid to how these changing currents were modeled within \code{ThinCurr}. 

\par
\begin{figure}[H]
\begin{center}
    \includegraphics[height=7cm]{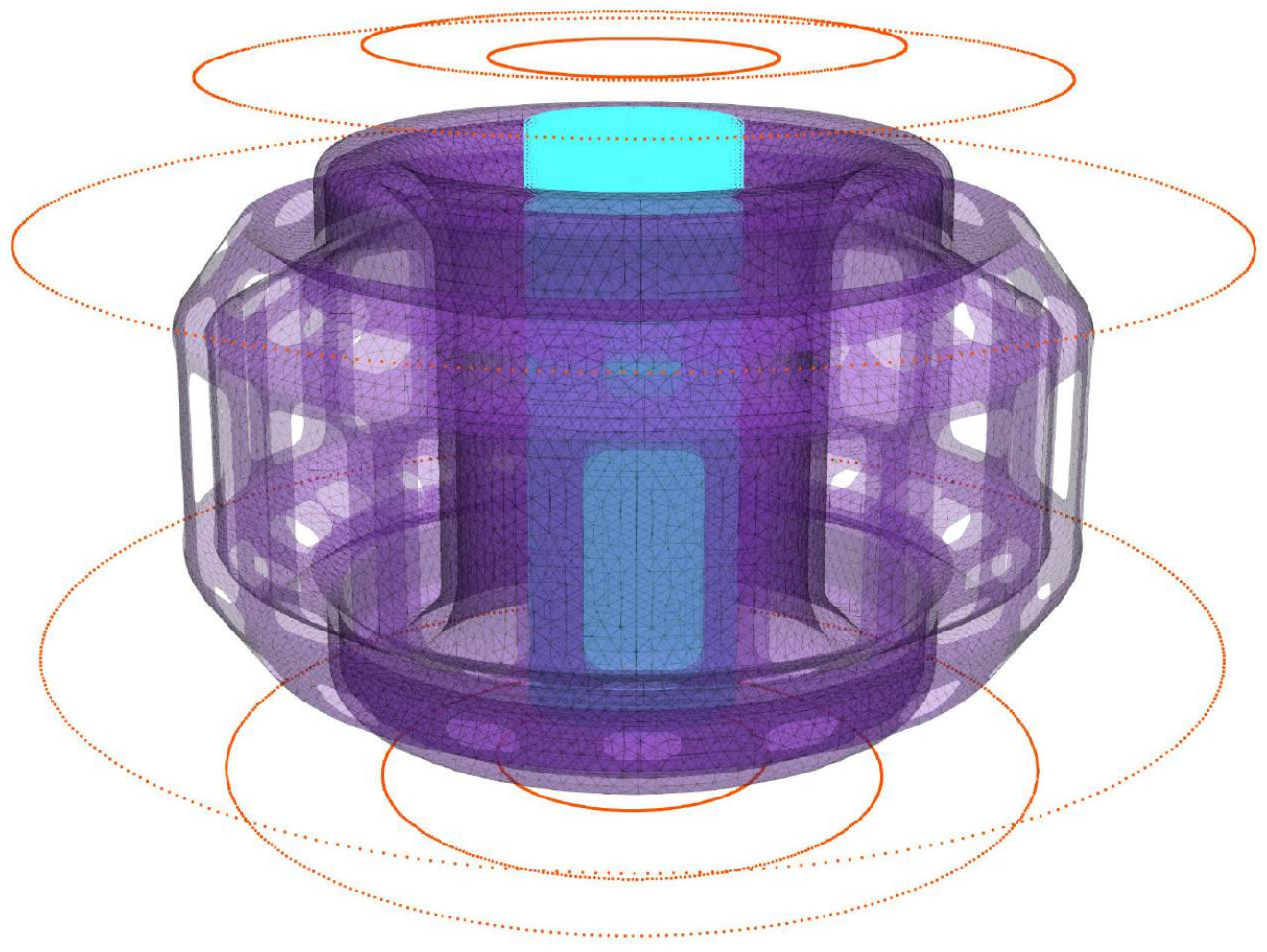} 
    \caption{Visualization of the \code{ThinCurr} model of SPARC including the vacuum vessel (purple), poloidal field coils (orange), and the central solenoid coils (blue).}
    \label{SPARCcoils}
\end{center}
\end{figure}

A \code{ThinCurr} model can be driven by prescribing either a voltage or current waveform to a set of predefined filaments. These filaments can be defined by a major radius and vertical position pair or by a collection of points in Cartesian space. The latter provides greater flexibility and the potential to define a coil with 3D features. The most straightforward application involves using a single filament to represent an external magnetic field coil as pictured in Figure~\ref{SPARCcoils}.  As implemented in the SPARC \code{ThinCurr} model for the central solenoid coils, driven filaments can also be grouped to represent coils with non-negligible height and width. In this case the resolution in both height and width is set by the user with the total coil current shared evenly across the filament set. 

\par
A more sophisticated driver configuration is used to represent currents flowing in the plasma which was used for all disruption related studies described in this article. In this configuration, circular filaments are laid out in an even rectangular (R-Z) grid with the desired resolution. This grid is them trimmed according to the size and shape of the last closed flux-surface from a reference plasma equilibrium. These filaments are then independently driven to recreate the desired current distribution. These current distributions typically correspond to to equilibrium reconstructions from historical discharges for the DIII-D cases and a target equilibrium for the SPARC cases. 

\begin{figure}[H]
\begin{center}
    \includegraphics[height=7cm]{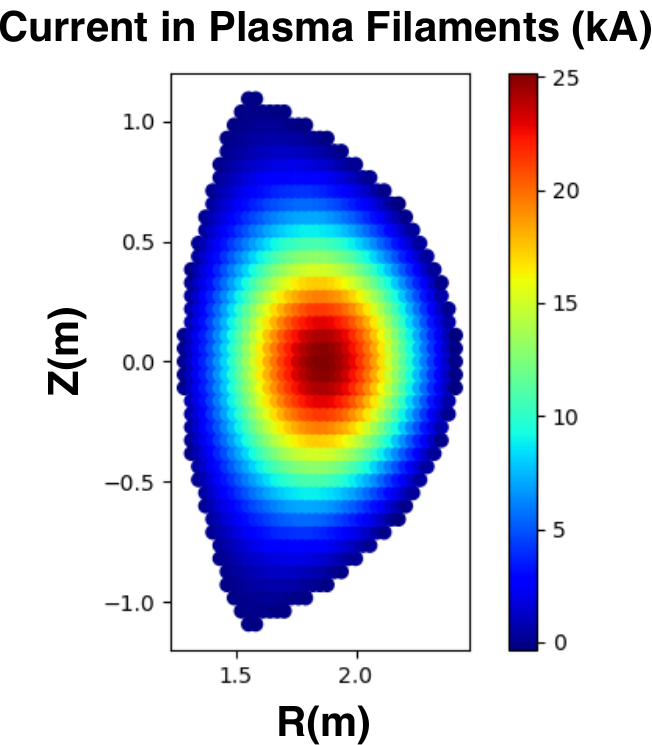} 
    \caption{Visualization of current flowing in filaments used to model a SPARC calculated equilibrium current profile in \code{ThinCurr}. Each colored point corresponds to a different filament in the model with a resolution designed to match that for the equilibrium reconstruction (16.2cm$^2$).}
    \label{curProfileDriver}
\end{center}
\end{figure}

\par
The currents in the filaments can then be scaled down to simulate a plasma current quench. While the current is scaled the profile can be held rigid or a series of profiles can be provided to allow the structure of the currents to change with time. An example of this technique is shown in Figure \ref{curProfileDriver} for a typical SPARC equilibrium. Note that due to the peakedness of this profile there is significantly less coupling to the wall than a flat profile with identical plasma current and significantly more coupling than using a single filament at the equilibrium current centroid. 

\par
Initially the REMC coils were modeled as a single three-dimensional filament; however, in order to capture the effects of the rectangular coil cross-section the coil was instead modeled as a sheet and meshed with a 10cm resolution. This allowed for the cross-sectional height to be accurately captured, and the thickness of the coil is then accounted for by modifying the modeled resistivity. These design choices enabled the coil geometry to accurately match the latest versions of both the SPARC and DIII-D REMCs.

\subsection{Effect of Coil Standoff Height - High-Field Side Coil on DIII-D}

\par
One consequence of placing the DIII-D coil on the high-field side is that it will be located very near the centerpost vacuum vessel and will therefore experience significant screening from induced eddy currents. As this effect is expected to depend sensitively on the distance between the coil and these conductors, a study was completed on the effect of the height of the coil's standoff ``feet" which will be used to offset the coil from the vessel. Note that both the coil and desired standoff feet will need to fit underneath the plasma facing tiles, and there is therefore an upper limit on the largest possible standoff of approximately 4cm.  

\par
\begin{figure}[H]
\begin{center}
    \includegraphics[height=9cm]{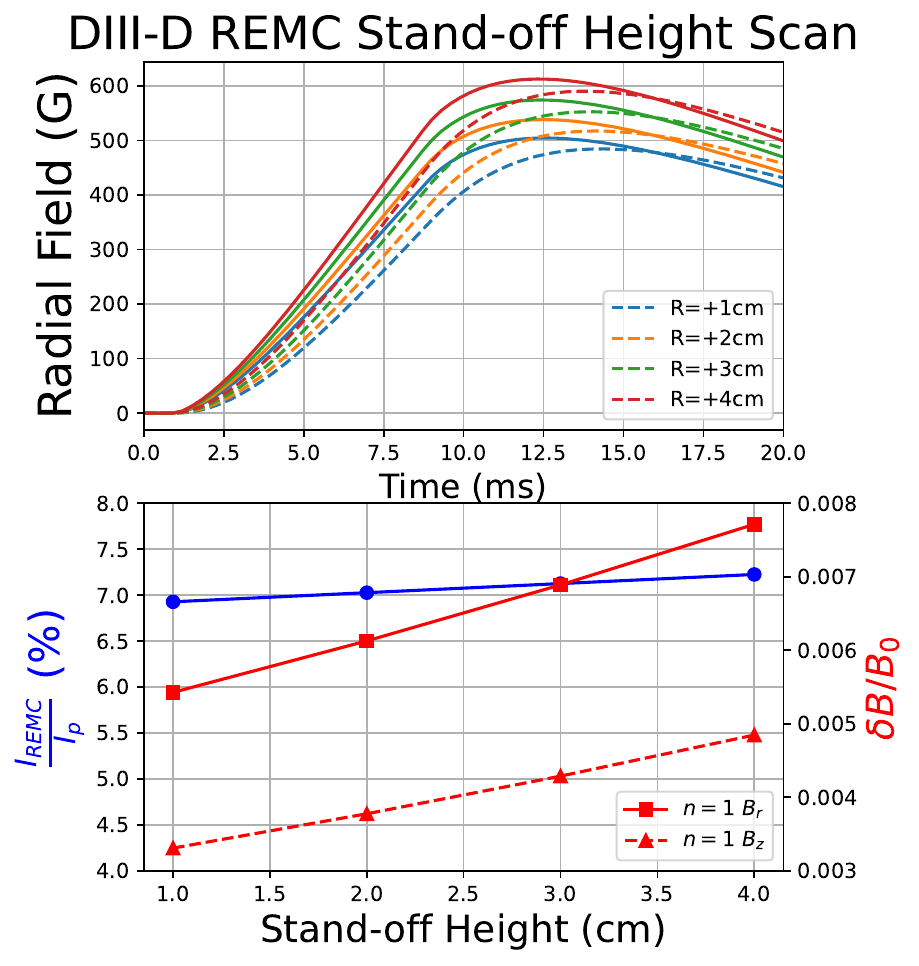} 
    \caption{A summary of a study which scanned the distance between the REMC and DIII-D centerpost. The top panel shows the $n=1$ radial field at the high-field side midplane of the plasma as a function of time with (dashed) and without (solid) eddy current screening. The bottom panel shows the maximum coupling efficiency (blue) and the radial (square) and vertical (triangle) $n=1$ fields at the midpoint of the CQ (t=5~ms). These fields are normalized to the toroidal field at the midplane which is $2.2T$ in this case. }
    \label{StandoffScan}
\end{center}
\end{figure}

This standoff height was scanned for integer values from one to four centimeters with the results summarized in Figure \ref{StandoffScan}. In the top panel of this figure, the n=1 field at the plasma high-field side midplane is shown both with (dashed) and without (solid) screening-current effects. One consequence of a larger standoff height is that the coil is physically closer to the decaying plasma currents and therefore couples to a larger percentage of the pre-disruption current. This can be seen in both the coupling efficiency in Figure \ref{StandoffScan}(bottom) as well as the unscreened field which not only benefits from the increased coil current but also the coil being nearer the measurement location (shown in Figure \ref{DIIID_fields}) at the plasma edge. This improved coupling results in a 4.1\% increase in the induced current; however, the true parameter of interest is not induced current but the applied magnetic field early in the current quench. This parameter is significantly affected by the inclusion of screening effects which do not significantly affect the maximum field applied by the coil but instead shift this maximum later in time. For simplicity these fields are evaluated at the midpoint of the CQ, which for the simulations presented in Figure \ref{StandoffScan} corresponds to t=5ms. At this point in the discharge the combination of large screening currents and the coil being physically closer to the plasma results in the 29.6\% difference for the radial field and a 31.7\% difference of the vertical field between the 1cm and 4cm case. It is therefore recommended that the DIII-D coil be designed with the largest possible standoff allowed by the in-vessel graphite tiles in order to maximize both coupled current and applied non-axisymmetric field. 

\subsection{Effect of Coil Resistance}
A study was then completed to determine the effect of the total coil resistance which, as mentioned previously, will be set in experiment using a variable resistor selected by the operator. This was modeled in \code{ThinCurr} by altering the resistivity of the horizontal connection leg (outside of vacuum vessel) of the coil while the majority of the coil was modeled with a resistance corresponding to height and width of the most up-to-date conductor design. This resistance value was varied from 0.1~mOhm to 100~mOhm for the DIII-D model and 0.1~mOhm to 500~mOhm for the SPARC model. These ranges were picked based on the desire to identify two distinct regimes for coil operation. The first regime will attempt to keep the coil current low for early commissioning and model validation. Once confidence is gained in the design the second regime will attempt to maximize the applied $n=1$ field while still staying below a maximum allowable current set by structural loading limits.

\par
\begin{figure}[H]
\begin{center}
    \includegraphics[height=7cm]{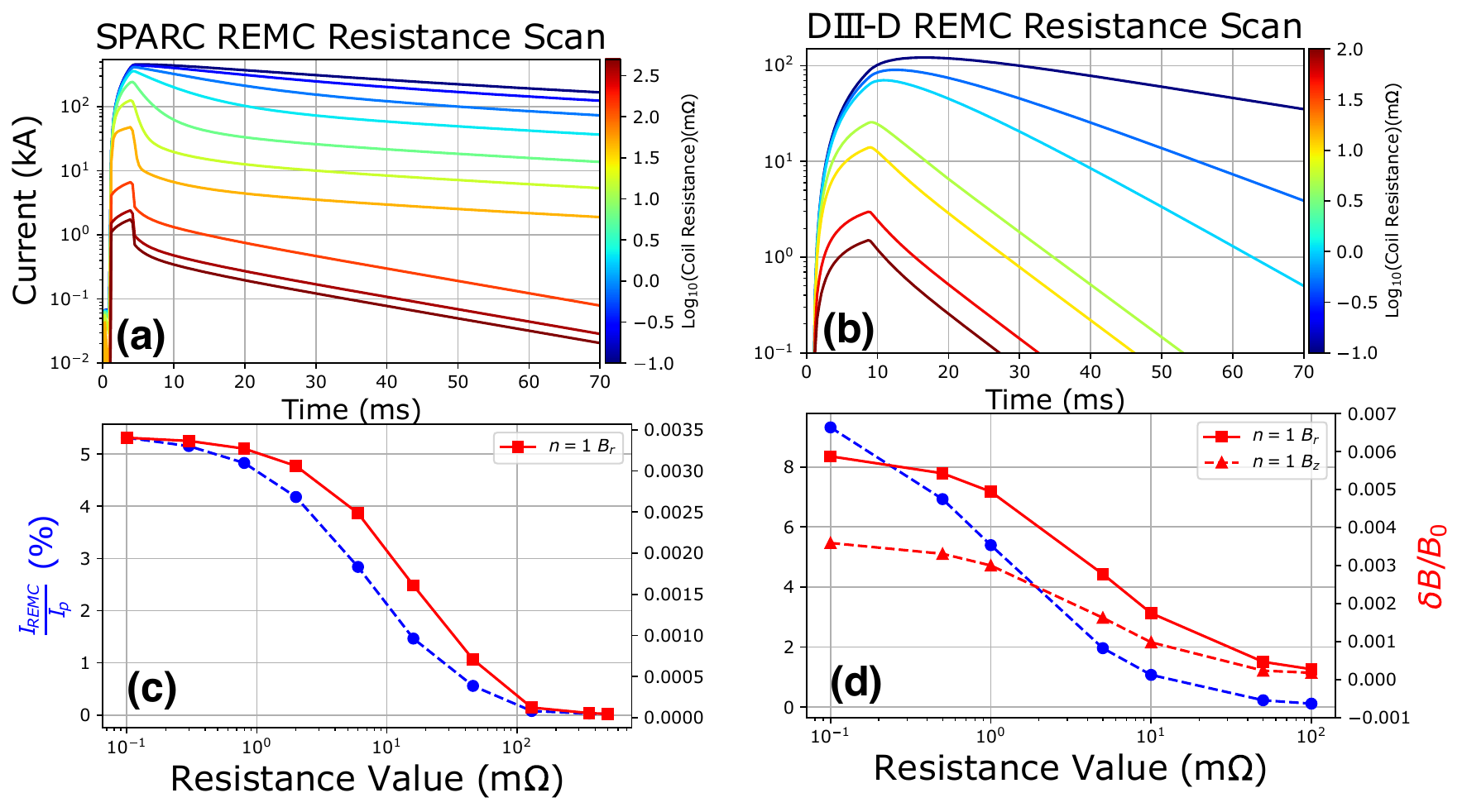} 
    \caption{A summary of a study of the effect of coil resistance on the performance of the SPARC and DIII-D REMC. The top plot shows the coil currents as a function of time. The bottom panel shows the coupling efficiency (blue) and the $n=1$ radial (square) and vertical (triangle) fields at the midpoint of the CQ normalized to the field on axis ($12.2T$ for SPARC and $2.2T$ for DIII-D). Since the vertical field is always near zero for the SPARC coil, with the measurement location shown in Figure \ref{DIIID_fields}, it is omitted.}
    \label{Rscan}
\end{center}
\end{figure}

As expected, the resistance was found to affect both the magnitude and time-dependence of the induced current in the DIII-D REMC. The results of this study including the coupling efficiency and fields applied at the midpoint of the CQ are summarized in Figures \ref{Rscan}(b,d). The variable resistor is shown to allow the induced current to be reduced by several orders of magnitude which will ensure access to both coil operating regimes. 

A similar study of the coil resistance was completed for the proposed SPARC REMC and is summarized in Figures \ref{Rscan}(a,c). Note that the SPARC REMC applies a much smaller (near zero) vertical field at the vessel midplane and therefore only the radial field varies significantly in this scan. This near zero field is a consequence of cancellation between the fields generated by top and bottom legs of the coil. Due to the significantly higher predisruption plasma current in SPARC, the coil resistance was varied over a slightly larger range in this scan from 0.1~mOhm to 500~mOhm and, similar to the DIII-D case, allows the induced coil current to vary over several orders of magnitude. This large range will allow for distinct commissioning and RE avoidance regimes to be established for this coil and allows hundreds of Gauss to be applied at the plasma early in the CQ when needed \cite{izzo_runaway_2022}.

\subsection{Effect of Current Quench Duration}
\par
Studies were then conducting to evaluate how robust each REMC would be to varying disruption conditions. The first case was motivated by uncertainty of the duration of a typical current quench in SPARC. Since the SPARC tokamak will operate in a unique plasma physics regime where MA runaway beams are expected, an effective and robust REMC is desired to reduce the number of melt events. Therefore, the effect of the CQ duration on the coil performance parameters was explored using the \code{ThinCurr} code. While a well established scaling law allows for confident estimations of the shortest CQ duration, the largest possible values for SPARC are unknown. Therefore, for the study presented here, these values were scanned from the estimated shortest CQ time of $\tau_{CQ} = 3.2$ms to a significantly higher upper limit of $\tau_{CQ}=50$ms \cite{sweeney_mhd_2020}.

\begin{figure}[H]
\begin{center}
    \includegraphics[height=6cm]{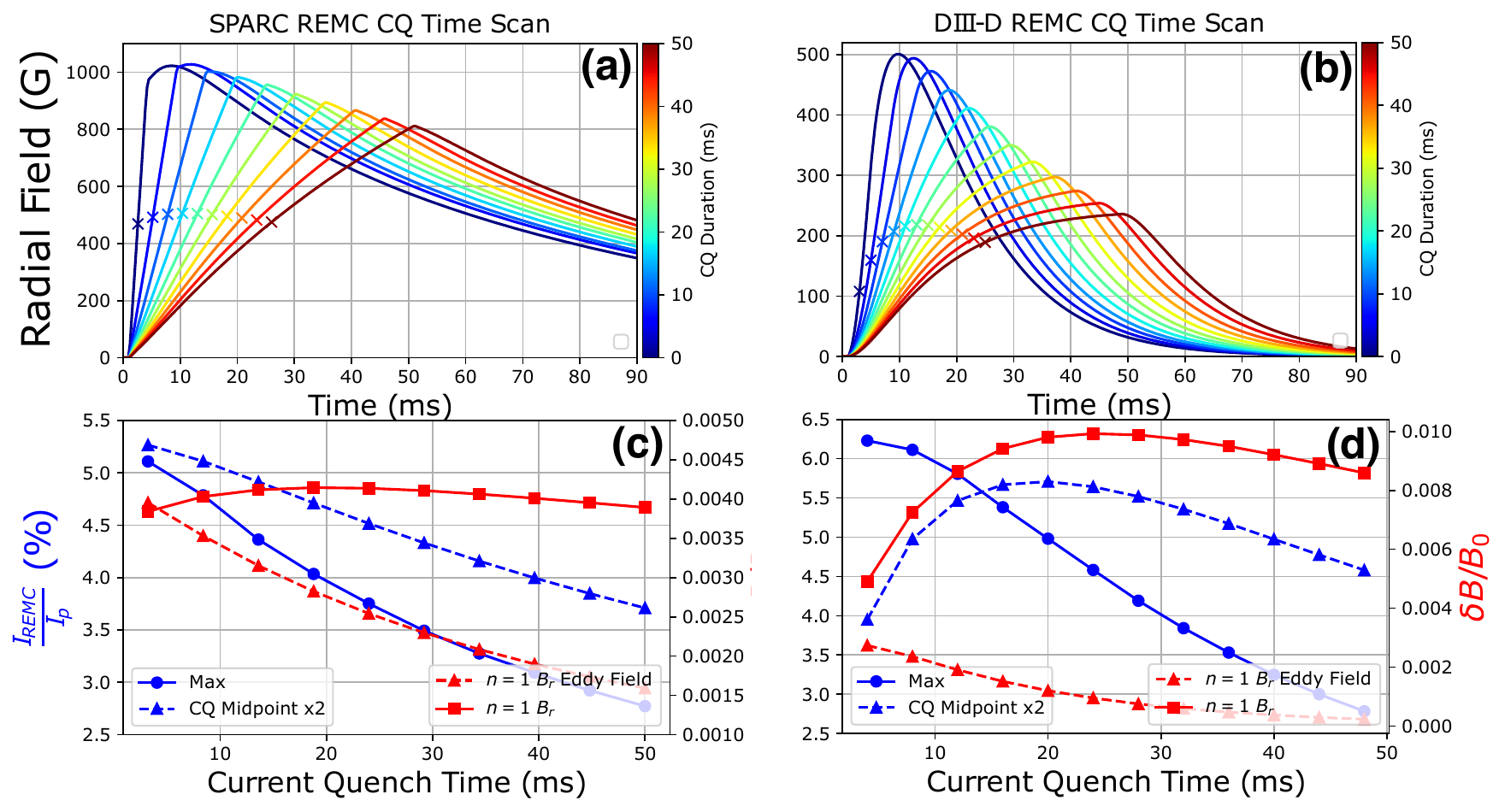} 
    \caption{A summary of a study of the effect of current quench duration on the performance of the SPARC (a,c) and DIII-D (b,d) REMC. The top plots show the $n=1$ radial field at the plasma surface where the colored `X' indicates the halfway point of the current quench. The bottom panel blue points correspond to the coupling efficiency at the maximum value (circles) and at the CQ midpoint (triangles) multiplied by two to allow for a direct comparison. The red points correspond to $n=1$ radial field at the CQ midpoint from the REMC (squares) and eddy currents (triangles) normalized to the field on axis ($12.2T$ for SPARC and $2.2T$ for DIII-D) }
    \label{CQscan}
\end{center}
\end{figure}

\par
In this study the current profile corresponds to that of a typical 8.7MA SPARC plasma, and this static profile is uniformly decreased linearly with time with a rate corresponding to the $\tau_{CQ}$ of interest. A summary of this study is presented in Figures \ref{CQscan}(a,c) where the time dependent radial fields, measured at the low-field side midplane, are shown along with the maximum coupled current and applied radial field halfway through the CQ. As expected, the maximum current induced in the coil decreases with CQ duration due to reduced toroidal loop voltage. However, the cases with shorter CQ durations and therefore higher loop voltages also drive significantly larger eddy currents which at the midpoint of the current quench can be quite significant as indicated in Figure \ref{CQscan}(c) using the red triangles. Similar to the coil fields, these radial eddy current fields are evaluated as the midpoint of each current quench time. These increased screening currents cancel the field driven by larger currents for faster CQ cases and result in an applied field which is a weak function of CQ time. Therefore, while the total induced current varied by up to $43\%$ in this scan the radial field applied to the plasma varied by less than $5\%$. This is an unexpected but comforting result that the REMC will be capable of applying large non-axisymmetric fields at the midpoint of the CQ regardless of the details of the current quench duration for the range of times explored. Also note that while the applied radial fields exceeds 800G for all CQ cases explored only around 500G is applied at this crucial stage.

\par
While the duration of CQs on DIII-D is much better understood, the effect of CQ times on the performance of the DIII-D REMC was also explored using the \code{ThinCurr} code. This study will allow for further model validation in early experiments implementing the REMC on DIII-D. The results of this study including the time-dependent fields, coupling efficiency, and the coil and eddy current radial fields at $t=\tau_{CQ}/2$ are shown in Figures \ref{CQscan}(b,d). One immediate observation is, while CQs were modeled using a linear ramp-down just like the SPARC cases, the time-dependent fields are significantly different shapes due to the proximity of the REMC to the centerpost. Similar to the SPARC cases, the applied radial field has a somewhat unexpected trend. While the coupling efficiency is a quickly decreasing function of $\tau_{CQ}$, the applied radial field has a local maximum near $\tau_{CQ}= 25ms$ with the fastest CQ ($\tau_{CQ}= 4ms$) actually resulting in the lowest radial field for the cases explored.

\subsection{Effect of Plasma Vertical Position}
\par
\begin{figure}[H]
\begin{center}
    \includegraphics[height=6cm]{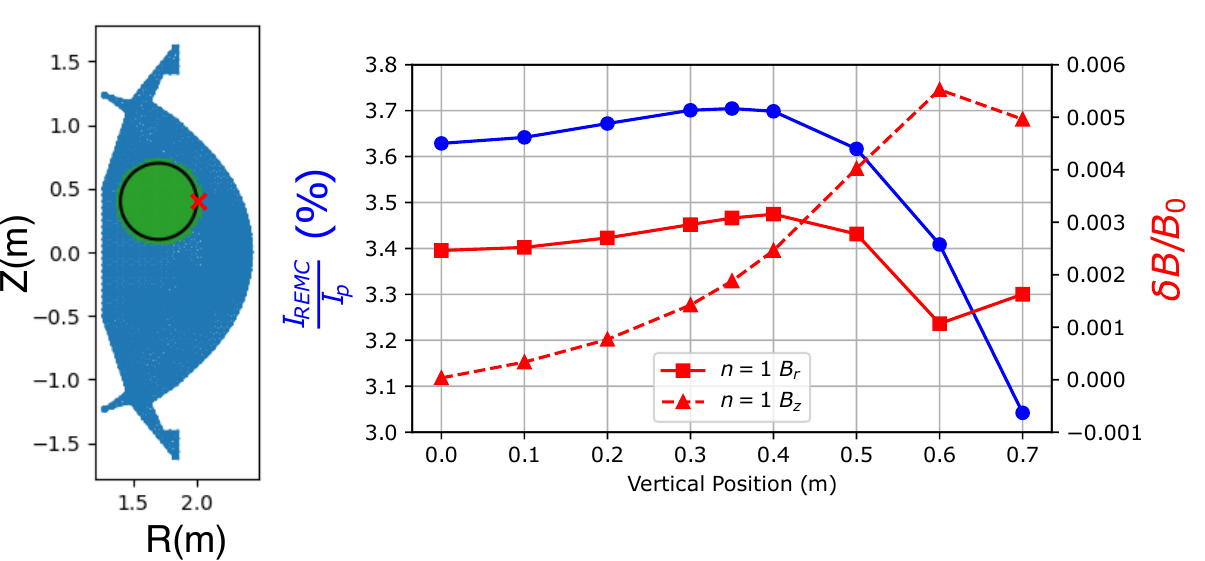} 
    \caption{A summary of the effect of plasma vertical position on the performance of the SPARC REMC. The left figure shows an example of a shifted equilibrium ($Z=0.4m$) with the measurement location indicated with the red `X'. The panel on the right shows the coupling efficiency (blue) and sampled $n=1$ vertical (triangle) and radial (square) fields as a function of vertical position. These fields are normalized to the field on axis ($12.2T$).}
    \label{SparcZscan}
\end{center}
\end{figure}
\begin{figure}[H]
\begin{center}
    \includegraphics[height=6cm]{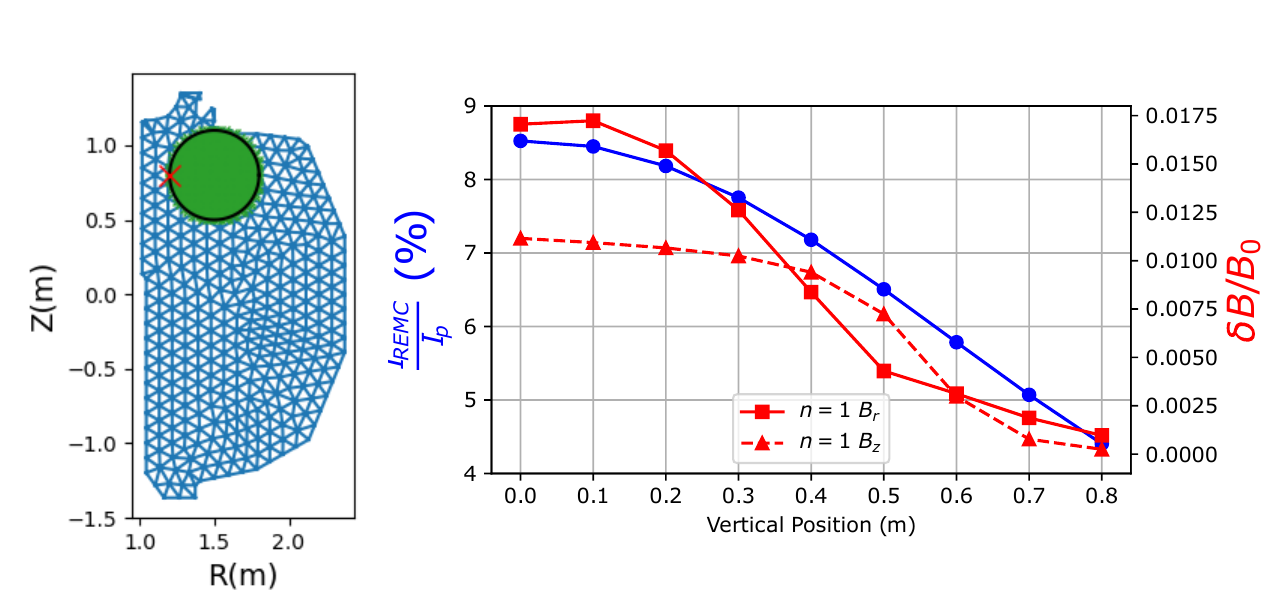} 
    \caption{A summary of the effect of plasma vertical position on the performance of the DIII-D REMC. The left figure shows an example of a shifted equilibrium with the measurement location indicated with the red `X'. The panel on the right shows the coupling efficiency (blue) and sampled vertical (triangle) and radial (square) fields as a function of vertical position. }
    \label{DIIIDZscan}
\end{center}
\end{figure}

Another potential scenario was the possibility of a “hot” vertical displacement event (VDE) which would result in the CQ occurring in a position vertically away from the mid-plane \cite{xue_hot_2019, nakamura_p-collapse-induced_1996, nakamura_mechanism_1996, lukash_influence_2007, gruber_vertical_1993,fitzpatrick_simple_2009, artola_3d_2021, eidietis_diffusive_2011, humphreys_analytic_1999}. Since both the SPARC and DIII-D vacuum vessels are up-down symmetric, the focus of this study was narrowed to consider only upward VDEs, and the plasma cross-section was chosen to be a circle for simplicity. This scenario was modeled assuming that the plasma current does not change significantly during the VDE. Therefore, the CQ can be modeled using a stationary collection of current carrying filaments at the final vertical position with the largest shift corresponding to the highest possible shift without reducing the plasma cross-section. This final position was scanned and the non-axisymmetric fields were measured at the plasma midplane on the side closest to the coil. The measurement locations are indicated using a red `X' in Figures \ref{SparcZscan} and \ref{DIIIDZscan}.

\par
The results of the SPARC coil study are highlighted in Figure \ref{SparcZscan}. As expected, the change in vertical position affects the coils ability to inductively couple to the changing plasma current; however, a bit unexpectedly the maximum coupling does not occur at the center of the vessel and instead occurs near a vertical shift of $\Delta Z$ = 35cm before falling off. Overall this change in coupling is a minor effect with a maximum difference of $18\%$ in the induced current. This reduction of current for vertical shifts above $\Delta Z = 35$cm is more than made up for by the addition of n=1 vertical field which grows as the plasma approaches the upper leg of the coil. In fact at the same point as a reduction is seen in the induced current the applied vertical field becomes larger than radial field (exceeds 600G) as the plasma approaches the top of the vessel. Also note that while the vertical motion of the plasma during the VDE will induce a toroidal loop voltage this voltage is not expected to be significant enough to close the switch on the SPARC REMC.

\par
The results of a similar study completed for the DIII-D REMC are presented in Figure \ref{DIIIDZscan}. Similar to the CQ duration study, these results will serve as a useful comparison to potential DIII-D experiments following the installation of the coil which would assess the coil's performance and robustness as well as serve as a unique opportunity for model validation. Due to the location of the DIII-D coil at the midplane of the DIII-D vacuum vessel centerpost, it experiences a steep drop-off in performance as the CQ occurs at higher vertical positions. This can be seen in the coils, coupling efficiency which drops from $8.52\%$ to $4.40\%$ when the CQ occurs in the top of the vessel. Unlike the SPARC REMC which has the added benefit of a larger $n=1$ vertical field for higher locations, the DIII-D coil suffers from a two-fold reduction. Not only is there less current induced in the coil, but the plasma is also physically further away which results in both the radial and vertical $n=1$ fields being reduced to less than 25G for the highest shifted case. At these fields the coil is not expected to be capable of significantly affecting the runaway electron beam. Therefore, these studies indicate that the low-field side coil has a clear advantage in terms of robustness to hot VDEs.

\section{REMC and Vacuum Vessel Forces}
\label{sec:Forces}

\par
\begin{figure}[H]
\begin{center}
    \includegraphics[height=9cm]{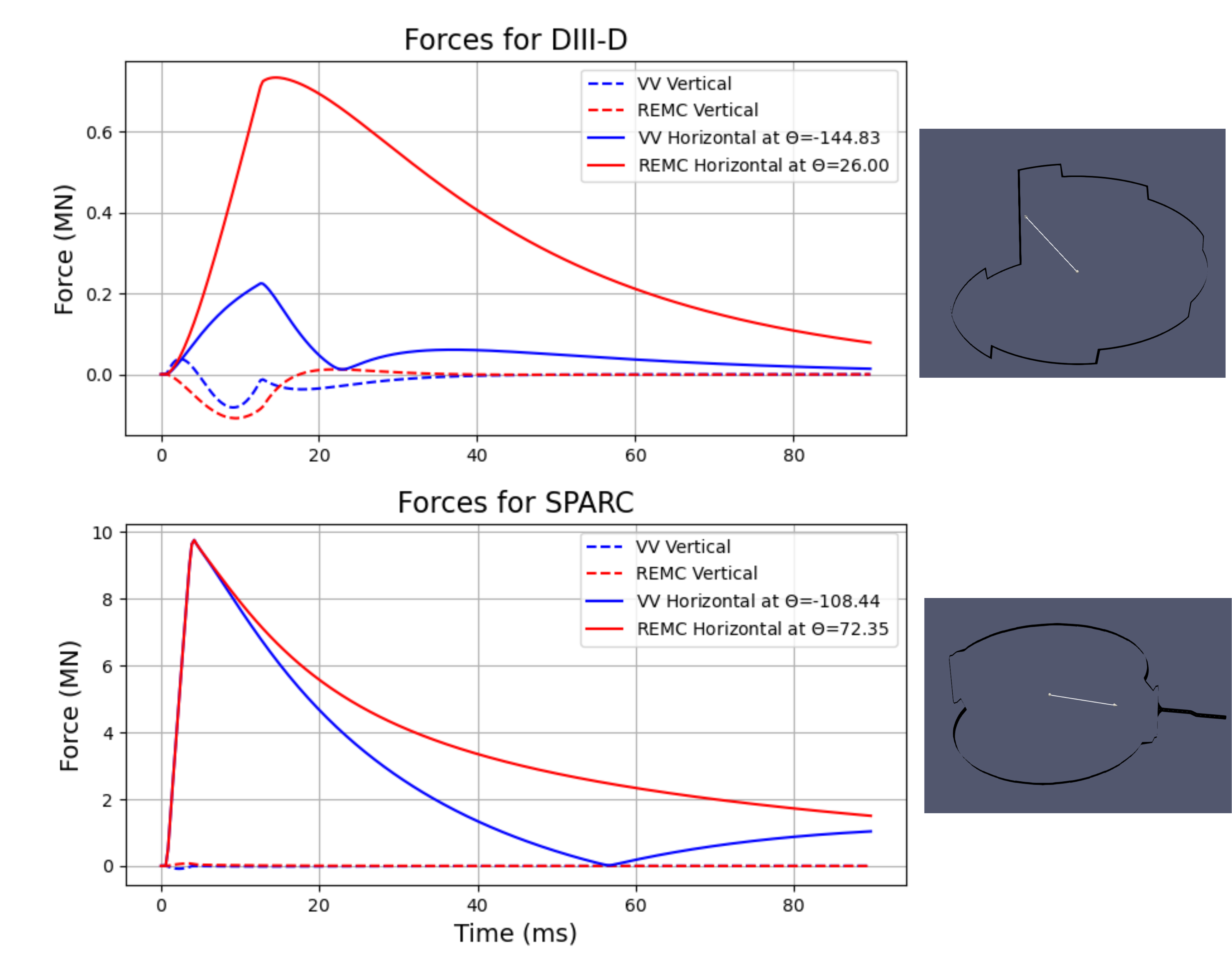} 
    \caption{The disruption-induced forces for both the DIII-D and SPARC tokamaks. Shown for both the vacuum vessel due to eddy currents (blue) and REMC (red) with the vertical force shown as a dotted line and horizontal forces shown as a solid line. The horizontal force vector is also shown relative to the REMC in the bottom right. }
    \label{REMC_forces}
\end{center}
\end{figure}

Due to \code{ThinCurr}'s ability to calculate both the eddy current induced in conducting structures and the magnetic fields produced from external coil sets in a fully 3D geometry it is uniquely suited to calculate the $\Vec{j} \cross \Vec{B}$ forces resulting from tokamak disruptions. This is of particular interest for REMCs which will conduct 100s of kAs of current in the presence of large toroidal magnetic fields. The results of these calculations for the DIII-D REMC  and vacuum vessel are shown in Figure \ref{REMC_forces}(top), respectively. As expected, the largest forces induced in the horizontal plane are a result of currents flowing in the longest vertical leg of the coil where the currents are orthogonal to the toroidal magnetic field. This is evident by the angle of the force which points directly at this segment of the coil. Interestingly the largest force on the vacuum vessel during a disruption results from the eddy currents formed in response to the REMC. This force points in the opposite direction to the REMC force but with a significantly lower magnitude. With less than a $1 MN$ of force on both the REMC and vacuum vessel these transient forces should be manageable when properly accounted for. 

\par
Due to the significantly larger toroidal field and plasma current in SPARC, it is crucial to accurately predict the forces induced by the REMC and its resulting eddy currents in order to ensure its safe operation. Similar to the DIII-D study the forces calculated by \code{ThinCurr} are shown in Figure \ref{REMC_forces}(bottom) for both the REMC and vacuum vessel. These forces exceed those predicted for DIII-D by an order of magnitude with forces transiently approaching 10MN. Once again these forces are pointed towards the vertical legs where current flows orthogonal to the magnetic field. The toroidal legs of the coil also interact with the poloidal field (PF) coils, producing a net overturning moment on the vacuum vessel. This will be explored in a future work. An interesting difference from the DIII-D study is that the vessel eddy current and REMC forces are nearly equal and opposite, and since the REMC is supported by the vessel, the net force on the vessel is greatly reduced. This is most likely due to the substantially thicker and therefore more conducting vacuum vessel where again these forces are a result of eddy currents in response to the REMC fields. The magnitude and direction of these forces as well as any induced torques have to be carefully considered when designing the mounting structures for the coil.

\section{Conclusion}
\label{sec:Conclusion}

The ability to accurately model currents induced in passive conducting structures is a crucial aspect of tokamak design. A newly developed, finite-element, electromagnetic inductive model \code{ThinCurr} has been leveraged for disruption studies for both the SPARC and DIII-D tokamaks. A highly detailed 3D model was developed for both SPARC and DIII-D within \code{ThinCurr} including varying material thickness, vessel ports, mounts for the REMC and VSCs (SPARC), and a passive REMC. The fundamental current decay time-scales in the vacuum vessel were found to vary significantly between the fully three-dimensional model and the simplified modeled created by sweeping the 2D cross-section. This further motivated the choice of fully three-dimensional analysis. 

\par
Particular attention was then placed on analyzing the effect of various parameters on the performance of the proposed REMC coils for both the SPARC and DIII-D devices. As a consequence of early design choices, this comparison serves as a unique study on the effect of coil location on several key performance parameters with the SPARC coil to be installed on the low-field side and the DIII-D coil on the high-field side. One consequence of the DIII-D REMC's location on the centerpost is that it experiences significant eddy current screening which reduces its ability to apply large fields during the early half of the current quench. This is not expected to limit its effectiveness, but it will be important to offset the coil as much as is allowed by the DIII-D plasma facing tiles. 

\par
The performance of the coils was then evaluated as a function of coil resistance which will be set in practice using an external resistor. The range of resistances explored indicate that it will be possible to both reduce the current for the first commissioning phase and maximize the applied fields for RE mitigation and prevention. The robustness of the coils were then evaluated for the case of varying current quench duration and plasma vertical position. Both REMCs were shown to robustly apply significant fields at the midpoint of the CQ regardless of the $\tau_{CQ}$ with the somewhat surprising result of slightly larger fields at moderate CQ durations. 

\par
The robustness of both coils was also assessed for the case of a hot VDE which would result in the CQ occurring away from the midplane. For the low-field side SPARC coil it was determined that any reduction in coupling efficiency was more than compensated by an increase in $n=1$ vertical field. The high-field side DIII-D coil showed a much less encouraging result as both the ability to couple to the predisruption plasma current and drive $n=1$ vertical and radial field quickly diminished as a function of vertical position. This shows a distinct advantage for the low-field side coil in dealing with these types of VDEs. 

\par
\code{ThinCurr} was then used to calculate the forces induced on both the vacuum vessel and REMC during a plasma current disruption. For both coils this force is a primary consequence of current flowing in the vertical segments of the coil, while the forces on the vessels are a consequence of eddy currents induced in response to the REMC fields. The local forces reacted between the REMC mounts and the vessel were calculated to be less than 1MN for DIII-D and less than 10MN for SPARC. The net body force on the vessel is the sum of the coil force and the vessel eddy current force and these simulations find that it is much reduced relative to the local forces. Engineering efforts are underway to ensure these forces can be tolerated in both devices.

\section{Acknowledgments}
The work presented in this article was supported through a combination of public and private funds. The results pertaining to the DIII-D tokamak were supported by the U.S. Department of Energy, Office of Science, Office of Fusion Energy Sciences, using the DIII-D National Fusion Facility, a DOE Office of Science user facility, under Award(s) DE-SC0022270, DE-SC0019239, DE-SC0014264, and DE-FC02-04ER54698, while the SPARC studies were supported by Commonwealth Fusion Systems. Some coauthors were supported through a combination of funding sources. A.F. Battey and C. Paz-Soldan were supported by DE-SC0022270 for their work relating to the DIII-D device and Commonwealth Fusion Systems for the work on the SPARC machine. Similarly, R. Sweeney was supported by DE-SC0014264 for collaboration on the DIII-D studies and by Commonwealth Fusion Systems for the SPARC REMC studies and C. Hansen was supported by DE-SC0019239 for work related to ThinCurr model development and DIII-D studies and by Commonwelath Fusion Systems for all SPARC research. D. Garnier, R.A.Tinguely, and A.J. Creely were supported entirely by Commonwealth Fusion systems while D. Weisberg was supported entirely through DE-FC02-04ER54698. 

\vspace{16pt}
\small
Disclaimer: This report was prepared as an account of work sponsored by an agency of the United States Government. Neither the United States Government nor any agency thereof, nor any of their employees, makes any warranty, express or implied, or assumes any legal liability or responsibility for the accuracy, completeness, or usefulness of any information, apparatus, product, or process disclosed, or represents that its use would not infringe privately owned rights. Reference herein to any specific commercial product, process, or service by trade name, trademark, manufacturer, or otherwise does not necessarily constitute or imply its endorsement, recommendation, or favoring by the United States Government or any agency thereof. The views and opinions of authors expressed herein do not necessarily state or reflect those of the United States Government or any agency thereof.

\vspace{20pt}
\printbibliography

@article{dreicer_electron_1959,
	title = {Electron and Ion Runaway in a Fully Ionized Gas. I},
	volume = {115},
	url = {https://link.aps.org/doi/10.1103/PhysRev.115.238},
	doi = {10.1103/PhysRev.115.238},
	pages = {238--249},
	number = {2},
	journaltitle = {Physical Review},
	shortjournal = {Phys. Rev.},
	author = {Dreicer, H.},
	urldate = {2023-01-18},
	date = {1959-07-15},
	note = {Publisher: American Physical Society},
}

@article{rosenbluth_theory_1997,
	title = {Theory for avalanche of runaway electrons in tokamaks},
	volume = {37},
	issn = {0029-5515},
	url = {https://dx.doi.org/10.1088/0029-5515/37/10/I03},
	doi = {10.1088/0029-5515/37/10/I03},
	pages = {1355},
	number = {10},
	journaltitle = {Nuclear Fusion},
	shortjournal = {Nucl. Fusion},
	author = {Rosenbluth, M. N. and Putvinski, S. V.},
	urldate = {2023-01-18},
	date = {1997-10},
	langid = {english},
}

@article{dreicer_electron_1960,
	title = {Electron and Ion Runaway in a Fully Ionized Gas. {II}},
	volume = {117},
	url = {https://link.aps.org/doi/10.1103/PhysRev.117.329},
	doi = {10.1103/PhysRev.117.329},
	pages = {329--342},
	number = {2},
	journaltitle = {Physical Review},
	shortjournal = {Phys. Rev.},
	author = {Dreicer, H.},
	urldate = {2023-01-18},
	date = {1960-01-15},
	note = {Publisher: American Physical Society},
}

@article{smith_hot_2008,
	title = {Hot tail runaway electron generation in tokamak disruptions},
	volume = {15},
	issn = {1070-664X},
	url = {https://aip-scitation-org.ezproxy.cul.columbia.edu/doi/10.1063/1.2949692},
	doi = {10.1063/1.2949692},
	pages = {072502},
	number = {7},
	journaltitle = {Physics of Plasmas},
	author = {Smith, H. M. and Verwichte, E.},
	urldate = {2023-01-18},
	date = {2008-07},
	note = {Publisher: American Institute of Physics},
}

@article{hanson_validation_2016,
	title = {Validation of conducting wall models using magnetic measurements},
	volume = {56},
	issn = {0029-5515},
	url = {https://dx.doi.org/10.1088/0029-5515/56/10/106022},
	doi = {10.1088/0029-5515/56/10/106022},
	pages = {106022},
	number = {10},
	journaltitle = {Nuclear Fusion},
	shortjournal = {Nucl. Fusion},
	author = {Hanson, J. M. and Bialek, J. and Turco, F. and King, J. and Navratil, G. A. and Strait, E. J. and Turnbull, A.},
	urldate = {2023-01-18},
	date = {2016-08},
	langid = {english},
	note = {Publisher: {IOP} Publishing},
}

@article{clement_optimal_2018,
	title = {optimal control techniques for resistive wall mode feedback in tokamaks},
	volume = {58},
	issn = {0029-5515},
	url = {https://dx.doi.org/10.1088/1741-4326/aaaecd},
	doi = {10.1088/1741-4326/aaaecd},
	pages = {046017},
	number = {4},
	journaltitle = {Nuclear Fusion},
	shortjournal = {Nucl. Fusion},
	author = {Clement, Mitchell and Hanson, Jeremy and Bialek, Jim and Navratil, Gerald},
	urldate = {2023-01-18},
	date = {2018-02},
	langid = {english},
	note = {Publisher: {IOP} Publishing},
}

@article{boozer_stabilization_1995,
	title = {Stabilization of resistive wall modes by slow plasma rotation},
	volume = {2},
	issn = {1070-664X},
	url = {https://aip-scitation-org.ezproxy.cul.columbia.edu/doi/10.1063/1.871009},
	doi = {10.1063/1.871009},
	pages = {4521--4532},
	number = {12},
	journaltitle = {Physics of Plasmas},
	author = {Boozer, Allen H.},
	urldate = {2023-01-18},
	date = {1995-12},
	note = {Publisher: American Institute of Physics},
}

@article{boozer_resistive_2003,
	title = {Resistive wall modes and error field amplification},
	volume = {10},
	issn = {1070-664X},
	url = {https://aip-scitation-org.ezproxy.cul.columbia.edu/doi/10.1063/1.1568751},
	doi = {10.1063/1.1568751},
	pages = {1458--1467},
	number = {5},
	journaltitle = {Physics of Plasmas},
	author = {Boozer, Allen H.},
	urldate = {2023-01-18},
	date = {2003-05},
	note = {Publisher: American Institute of Physics},
}

@article{boozer_robust_2004,
	title = {Robust feedback systems for resistive wall modes},
	volume = {11},
	issn = {1070-664X},
	url = {https://aip-scitation-org.ezproxy.cul.columbia.edu/doi/10.1063/1.1628687},
	doi = {10.1063/1.1628687},
	pages = {110--114},
	number = {1},
	journaltitle = {Physics of Plasmas},
	author = {Boozer, Allen H.},
	urldate = {2023-01-18},
	date = {2004-01},
	note = {Publisher: American Institute of Physics},
}

@article{boozer_simplified_2010,
	title = {Simplified multimode calculations of resistive wall modes},
	volume = {17},
	issn = {1070-664X},
	url = {https://aip-scitation-org.ezproxy.cul.columbia.edu/doi/10.1063/1.3453706},
	doi = {10.1063/1.3453706},
	pages = {072503},
	number = {7},
	journaltitle = {Physics of Plasmas},
	author = {Boozer, Allen H.},
	urldate = {2023-01-18},
	date = {2010-07},
	note = {Publisher: American Institute of Physics},
}

@article{clement_gpu-based_2017,
	title = {{GPU}-based optimal control for {RWM} feedback in tokamaks},
	volume = {68},
	issn = {0967-0661},
	url = {https://www.sciencedirect.com/science/article/pii/S0967066117301685},
	doi = {10.1016/j.conengprac.2017.08.002},
	pages = {15--22},
	journaltitle = {Control Engineering Practice},
	shortjournal = {Control Engineering Practice},
	author = {Clement, Mitchell and Hanson, Jeremy and Bialek, Jim and Navratil, Gerald},
	urldate = {2023-01-18},
	date = {2017-11-01},
	langid = {english},
}

@article{bialek_modeling_2001,
	title = {Modeling of active control of external magnetohydrodynamic instabilities},
	volume = {8},
	issn = {1070-664X},
	url = {https://aip-scitation-org.ezproxy.cul.columbia.edu/doi/10.1063/1.1362532},
	doi = {10.1063/1.1362532},
	pages = {2170--2180},
	number = {5},
	journaltitle = {Physics of Plasmas},
	author = {Bialek, James and Boozer, Allen H. and Mauel, M. E. and Navratil, G. A.},
	urldate = {2023-01-18},
	date = {2001-05},
	note = {Publisher: American Institute of Physics},
}

@article{katsuro-hopkins_enhanced_2007,
	title = {Enhanced {ITER} resistive wall mode feedback performance using optimal control techniques},
	volume = {47},
	issn = {0029-5515},
	url = {https://dx.doi.org/10.1088/0029-5515/47/9/012},
	doi = {10.1088/0029-5515/47/9/012},
	pages = {1157},
	number = {9},
	journaltitle = {Nuclear Fusion},
	shortjournal = {Nucl. Fusion},
	author = {Katsuro-Hopkins, O. and Bialek, J. and Maurer, D. A. and Navratil, G. A.},
	urldate = {2023-01-18},
	date = {2007-08},
	langid = {english},
}

@article{okabayashi_control_2005,
	title = {Control of the resistive wall mode with internal coils in the {DIII}–D tokamak},
	volume = {45},
	issn = {0029-5515},
	url = {https://dx.doi.org/10.1088/0029-5515/45/12/028},
	doi = {10.1088/0029-5515/45/12/028},
	pages = {1715},
	number = {12},
	journaltitle = {Nuclear Fusion},
	shortjournal = {Nucl. Fusion},
	author = {Okabayashi, M. and Bialek, J. and Bondeson, A. and Chance, M. S. and Chu, M. S. and Garofalo, A. M. and Hatcher, R. and In, Y. and Jackson, G. L. and Jayakumar, R. J. and Jensen, T. H. and Katsuro-Hopkins, O. and Haye, R. J. La and Liu, Y. Q. and Navratil, G. A. and Reimerdes, H. and Scoville, J. T. and Strait, E. J. and Takechi, M. and Turnbull, A. D. and Gohil, P. and Kim, J. S. and Makowski, M. A. and Manickam, J. and Menard, J.},
	urldate = {2023-01-18},
	date = {2005-11},
	langid = {english},
}

@article{mauel_dynamics_2005,
	title = {Dynamics and control of resistive wall modes with magnetic feedback control coils: experiment and theory},
	volume = {45},
	issn = {0029-5515},
	url = {https://dx.doi.org/10.1088/0029-5515/45/4/010},
	doi = {10.1088/0029-5515/45/4/010},
	shorttitle = {Dynamics and control of resistive wall modes with magnetic feedback control coils},
	pages = {285},
	number = {4},
	journaltitle = {Nuclear Fusion},
	shortjournal = {Nucl. Fusion},
	author = {Mauel, M. E. and Bialek, J. and Boozer, A. H. and Cates, C. and James, R. and Katsuro-Hopkins, O. and Klein, A. and Liu, Y. and Maurer, D. A. and Maslovsky, D. and Navratil, G. A. and Pedersen, T. S. and Shilov, M. and Stillits, N.},
	urldate = {2023-01-18},
	date = {2005-04},
	langid = {english},
}

@article{leuer_plasma_2010,
	title = {Plasma Startup Design of Fully Superconducting Tokamaks {EAST} and {KSTAR} With Implications for {ITER}},
	volume = {38},
	issn = {1939-9375},
	doi = {10.1109/TPS.2009.2037890},
	pages = {333--340},
	number = {3},
	journaltitle = {{IEEE} Transactions on Plasma Science},
	author = {Leuer, J. A. and Eidietis, N. W. and Ferron, J. R. and Humphreys, D. A. and Hyatt, A. W. and Jackson, G. L. and Johnson, R. D. and Penaflor, B. G. and Piglowski, D. A. and Walker, M. L. and Welander, A. S. and Yoon, S. W. and Hahn, S. H. and Oh, Y. K. and Xiao, B. J. and Wang, H. Z. and Yuan, Q. P. and Mueller, D.},
	date = {2010-03},
	note = {Conference Name: {IEEE} Transactions on Plasma Science},
}

@article{breizman_physics_2019,
	title = {Physics of runaway electrons in tokamaks},
	volume = {59},
	issn = {0029-5515},
	url = {https://dx.doi.org/10.1088/1741-4326/ab1822},
	doi = {10.1088/1741-4326/ab1822},
	pages = {083001},
	number = {8},
	journaltitle = {Nuclear Fusion},
	shortjournal = {Nucl. Fusion},
	author = {Breizman, Boris N. and Aleynikov, Pavel and Hollmann, Eric M. and Lehnen, Michael},
	urldate = {2023-01-25},
	date = {2019-06},
	langid = {english},
	note = {Publisher: {IOP} Publishing},
}

@article{connor_relativistic_1975,
	title = {Relativistic limitations on runaway electrons},
	volume = {15},
	issn = {0029-5515},
	url = {https://dx.doi.org/10.1088/0029-5515/15/3/007},
	doi = {10.1088/0029-5515/15/3/007},
	pages = {415},
	number = {3},
	journaltitle = {Nuclear Fusion},
	shortjournal = {Nucl. Fusion},
	author = {Connor, J. W. and Hastie, R. J.},
	urldate = {2023-02-07},
	date = {1975-06},
	langid = {english},
}

@article{aleynikov_theory_2015,
	title = {Theory of Two Threshold Fields for Relativistic Runaway Electrons},
	volume = {114},
	url = {https://link.aps.org/doi/10.1103/PhysRevLett.114.155001},
	doi = {10.1103/PhysRevLett.114.155001},
	pages = {155001},
	number = {15},
	journaltitle = {Physical Review Letters},
	shortjournal = {Phys. Rev. Lett.},
	author = {Aleynikov, Pavel and Breizman, Boris N.},
	urldate = {2023-02-09},
	date = {2015-04-14},
	note = {Publisher: American Physical Society},
}

@article{paz-soldan_kink_2019,
	title = {Kink instabilities of the post-disruption runaway electron beam at low safety factor},
	volume = {61},
	issn = {0741-3335},
	url = {https://dx.doi.org/10.1088/1361-6587/aafd15},
	doi = {10.1088/1361-6587/aafd15},
	pages = {054001},
	number = {5},
	journaltitle = {Plasma Physics and Controlled Fusion},
	shortjournal = {Plasma Phys. Control. Fusion},
	author = {Paz-Soldan, C. and Eidietis, N. W. and Liu, Y. Q. and Shiraki, D. and Boozer, A. H. and Hollmann, E. M. and Kim, C. C. and Lvovskiy, A.},
	urldate = {2023-04-28},
	date = {2019-03},
	langid = {english},
	note = {Publisher: {IOP} Publishing},
}

@article{reux_demonstration_2021,
	title = {Demonstration of Safe Termination of Megaampere Relativistic Electron Beams in Tokamaks},
	volume = {126},
	url = {https://link.aps.org/doi/10.1103/PhysRevLett.126.175001},
	doi = {10.1103/PhysRevLett.126.175001},
	pages = {175001},
	number = {17},
	journaltitle = {Physical Review Letters},
	shortjournal = {Phys. Rev. Lett.},
	author = {Reux, Cédric and Paz-Soldan, Carlos and Aleynikov, Pavel and Bandaru, Vinodh and Ficker, Ondrej and Silburn, Scott and Hoelzl, Matthias and Jachmich, Stefan and Eidietis, Nicholas and Lehnen, Michael and Sridhar, Sundaresan and {JET contributors}},
	urldate = {2023-04-28},
	date = {2021-04-30},
	note = {Publisher: American Physical Society},
}

@article{liu_mars-f_2019,
	title = {{MARS}-F modeling of post-disruption runaway beam loss by magnetohydrodynamic instabilities in {DIII}-D},
	volume = {59},
	issn = {0029-5515},
	url = {https://dx.doi.org/10.1088/1741-4326/ab3f87},
	doi = {10.1088/1741-4326/ab3f87},
	pages = {126021},
	number = {12},
	journaltitle = {Nuclear Fusion},
	shortjournal = {Nucl. Fusion},
	author = {Liu, Y. Q. and Parks, P. B. and Paz-Soldan, C. and Kim, C. and Lao, L. L.},
	urldate = {2023-04-28},
	date = {2019-10},
	langid = {english},
	note = {Publisher: {IOP} Publishing},
}

@article{paz-soldan_novel_2021,
	title = {A novel path to runaway electron mitigation via deuterium injection and current-driven {MHD} instability},
	volume = {61},
	issn = {0029-5515},
	url = {https://dx.doi.org/10.1088/1741-4326/ac2a69},
	doi = {10.1088/1741-4326/ac2a69},
	pages = {116058},
	number = {11},
	journaltitle = {Nuclear Fusion},
	shortjournal = {Nucl. Fusion},
	author = {Paz-Soldan, C. and Reux, C. and Aleynikova, K. and Aleynikov, P. and Bandaru, V. and Beidler, M. and Eidietis, N. and Liu, Y. Q. and Liu, C. and Lvovskiy, A. and Silburn, S. and Bardoczi, L. and Baylor, L. and Bykov, I. and Carnevale, D. and Negrete, D. Del-Castillo and Du, X. and Ficker, O. and Gerasimov, S. and Hoelzl, M. and Hollmann, E. and Jachmich, S. and Jardin, S. and Joffrin, E. and Lasnier, C. and Lehnen, M. and Macusova, E. and Manzanares, A. and Papp, G. and Pautasso, G. and Popovic, Z. and Rimini, F. and Shiraki, D. and Sommariva, C. and Spong, D. and Sridhar, S. and Szepesi, G. and Zhao, C. and Team, the {DIII}-D. and Contributors, J. E. T.},
	urldate = {2023-04-28},
	date = {2021-10},
	langid = {english},
	note = {Publisher: {IOP} Publishing},
}

@article{creely_overview_2020,
	title = {Overview of the {SPARC} tokamak},
	volume = {86},
	issn = {0022-3778, 1469-7807},
	url = {https://www.cambridge.org/core/journals/journal-of-plasma-physics/article/overview-of-the-sparc-tokamak/DD3C44ECD26F5EACC554811764EF9FF0},
	doi = {10.1017/S0022377820001257},
	pages = {865860502},
	number = {5},
	journaltitle = {Journal of Plasma Physics},
	author = {Creely, A. J. and Greenwald, M. J. and Ballinger, S. B. and Brunner, D. and Canik, J. and Doody, J. and Fülöp, T. and Garnier, D. T. and Granetz, R. and Gray, T. K. and Holland, C. and Howard, N. T. and Hughes, J. W. and Irby, J. H. and Izzo, V. A. and Kramer, G. J. and Kuang, A. Q. and {LaBombard}, B. and Lin, Y. and Lipschultz, B. and Logan, N. C. and Lore, J. D. and Marmar, E. S. and Montes, K. and Mumgaard, R. T. and Paz-Soldan, C. and Rea, C. and Reinke, M. L. and Rodriguez-Fernandez, P. and Särkimäki, K. and Sciortino, F. and Scott, S. D. and Snicker, A. and Snyder, P. B. and Sorbom, B. N. and Sweeney, R. and Tinguely, R. A. and Tolman, E. A. and Umansky, M. and Vallhagen, O. and Varje, J. and Whyte, D. G. and Wright, J. C. and Wukitch, S. J. and Zhu, J. and Team, the {SPARC}},
	urldate = {2023-05-04},
	date = {2020-10},
	langid = {english},
	note = {Publisher: Cambridge University Press},
}

@misc{m_lehnen_rd_2018,
	title = {R\&D {FOR} {RELIABLE} {DISRUPTION} {MITIGATION} {IN} {ITER}},
	url = {chrome-extension://efaidnbmnnnibpcajpcglclefindmkaj/https://nucleus.iaea.org/sites/fusionportal/Shared%20Documents/FEC%202018/fec2018-preprints/preprint0318.pdf},
	publisher = {Preprint: 2018 {IAEA} Fusion Energy Conf.},
	author = {{M. Lehnen} and {D.J. Campbell} and {D. Hu} and {U. Kruezi} and {T.C. Luce} and {S. Maruyama} and {J.A. Snipes} and {R. Sweeney} and {N.W. Eidietis} and {A. Matsuyama} and {E. Nardon}},
	date = {2018},
}

@article{hender_chapter_2007,
	title = {Chapter 3: {MHD} stability, operational limits and disruptions},
	volume = {47},
	issn = {0029-5515},
	url = {https://dx.doi.org/10.1088/0029-5515/47/6/S03},
	doi = {10.1088/0029-5515/47/6/S03},
	shorttitle = {Chapter 3},
	pages = {S128},
	number = {6},
	journaltitle = {Nuclear Fusion},
	shortjournal = {Nucl. Fusion},
	author = {Hender, T. C. and Wesley, J. C. and Bialek, J. and Bondeson, A. and Boozer, A. H. and Buttery, R. J. and Garofalo, A. and Goodman, T. P. and Granetz, R. S. and Gribov, Y. and Gruber, O. and Gryaznevich, M. and Giruzzi, G. and Günter, S. and Hayashi, N. and Helander, P. and Hegna, C. C. and Howell, D. F. and Humphreys, D. A. and Huysmans, G. T. A. and Hyatt, A. W. and Isayama, A. and Jardin, S. C. and Kawano, Y. and Kellman, A. and Kessel, C. and Koslowski, H. R. and Haye, R. J. La and Lazzaro, E. and Liu, Y. Q. and Lukash, V. and Manickam, J. and Medvedev, S. and Mertens, V. and Mirnov, S. V. and Nakamura, Y. and Navratil, G. and Okabayashi, M. and Ozeki, T. and Paccagnella, R. and Pautasso, G. and Porcelli, F. and Pustovitov, V. D. and Riccardo, V. and Sato, M. and Sauter, O. and Schaffer, M. J. and Shimada, M. and Sonato, P. and Strait, E. J. and Sugihara, M. and Takechi, M. and Turnbull, A. D. and Westerhof, E. and Whyte, D. G. and Yoshino, R. and Zohm, H. and the {ITPA} {MHD}, Disruption and Group, Magnetic Control Topical},
	urldate = {2023-05-08},
	date = {2007-06},
	langid = {english},
}

@article{lehnen_disruptions_2015,
	title = {Disruptions in {ITER} and strategies for their control and mitigation},
	volume = {463},
	issn = {0022-3115},
	url = {https://www.sciencedirect.com/science/article/pii/S0022311514007594},
	doi = {10.1016/j.jnucmat.2014.10.075},
	series = {{PLASMA}-{SURFACE} {INTERACTIONS} 21},
	pages = {39--48},
	journaltitle = {Journal of Nuclear Materials},
	shortjournal = {Journal of Nuclear Materials},
	author = {Lehnen, M. and Aleynikova, K. and Aleynikov, P. B. and Campbell, D. J. and Drewelow, P. and Eidietis, N. W. and Gasparyan, Yu. and Granetz, R. S. and Gribov, Y. and Hartmann, N. and Hollmann, E. M. and Izzo, V. A. and Jachmich, S. and Kim, S. -H. and Kočan, M. and Koslowski, H. R. and Kovalenko, D. and Kruezi, U. and Loarte, A. and Maruyama, S. and Matthews, G. F. and Parks, P. B. and Pautasso, G. and Pitts, R. A. and Reux, C. and Riccardo, V. and Roccella, R. and Snipes, J. A. and Thornton, A. J. and de Vries, P. C.},
	urldate = {2023-05-08},
	date = {2015-08-01},
	langid = {english},
}

@article{hollmann_status_2014,
	title = {Status of research toward the {ITER} disruption mitigation system},
	volume = {22},
	issn = {1070-664X},
	url = {https://doi.org/10.1063/1.4901251},
	doi = {10.1063/1.4901251},
	pages = {021802},
	number = {2},
	journaltitle = {Physics of Plasmas},
	shortjournal = {Physics of Plasmas},
	author = {Hollmann, E. M. and Aleynikov, P. B. and Fülöp, T. and Humphreys, D. A. and Izzo, V. A. and Lehnen, M. and Lukash, V. E. and Papp, G. and Pautasso, G. and Saint-Laurent, F. and Snipes, J. A.},
	urldate = {2023-05-08},
	date = {2014-11-17},
}

@article{boozer_theory_2015,
	title = {Theory of runaway electrons in {ITER}: Equations, important parameters, and implications for mitigation},
	volume = {22},
	issn = {1070-664X},
	url = {https://doi.org/10.1063/1.4913582},
	doi = {10.1063/1.4913582},
	shorttitle = {Theory of runaway electrons in {ITER}},
	pages = {032504},
	number = {3},
	journaltitle = {Physics of Plasmas},
	shortjournal = {Physics of Plasmas},
	author = {Boozer, Allen H.},
	urldate = {2023-05-08},
	date = {2015-03-10},
}

@article{reux_runaway_2015,
	title = {Runaway electron beam generation and mitigation during disruptions at {JET}-{ILW}},
	volume = {55},
	issn = {0029-5515},
	url = {https://dx.doi.org/10.1088/0029-5515/55/9/093013},
	doi = {10.1088/0029-5515/55/9/093013},
	pages = {093013},
	number = {9},
	journaltitle = {Nuclear Fusion},
	shortjournal = {Nucl. Fusion},
	author = {Reux, C. and Plyusnin, V. and Alper, B. and Alves, D. and Bazylev, B. and Belonohy, E. and Boboc, A. and Brezinsek, S. and Coffey, I. and Decker, J. and Drewelow, P. and Devaux, S. and Vries, P. C. de and Fil, A. and Gerasimov, S. and Giacomelli, L. and Jachmich, S. and Khilkevitch, E. M. and Kiptily, V. and Koslowski, R. and Kruezi, U. and Lehnen, M. and Lupelli, I. and Lomas, P. J. and Manzanares, A. and Aguilera, A. Martin De and Matthews, G. F. and Mlynář, J. and Nardon, E. and Nilsson, E. and Thun, C. Perez von and Riccardo, V. and Saint-Laurent, F. and Shevelev, A. E. and Sips, G. and Sozzi, C. and contributors, J. E. T.},
	urldate = {2023-05-08},
	date = {2015-08},
	langid = {english},
	note = {Publisher: {IOP} Publishing},
}

@article{papp_effect_2013,
	title = {The effect of {ITER}-like wall on runaway electron generation in {JET}},
	volume = {53},
	issn = {0029-5515},
	url = {https://dx.doi.org/10.1088/0029-5515/53/12/123017},
	doi = {10.1088/0029-5515/53/12/123017},
	pages = {123017},
	number = {12},
	journaltitle = {Nuclear Fusion},
	shortjournal = {Nucl. Fusion},
	author = {Papp, G. and Fülöp, T. and Fehér, T. and Vries, P. C. de and Riccardo, V. and Reux, C. and Lehnen, M. and Kiptily, V. and Plyusnin, V. V. and Alper, B. and contributors, {JET} {EFDA}},
	urldate = {2023-05-08},
	date = {2013-11},
	langid = {english},
	note = {Publisher: {IOP} Publishing and International Atomic Energy Agency},
}

@article{gobbin_runaway_2017,
	title = {Runaway electron mitigation by 3D fields in the {ASDEX}-Upgrade experiment},
	volume = {60},
	issn = {0741-3335},
	url = {https://dx.doi.org/10.1088/1361-6587/aa90c4},
	doi = {10.1088/1361-6587/aa90c4},
	pages = {014036},
	number = {1},
	journaltitle = {Plasma Physics and Controlled Fusion},
	shortjournal = {Plasma Phys. Control. Fusion},
	author = {Gobbin, M. and Li, L. and Liu, Y. Q. and Marrelli, L. and Nocente, M. and Papp, G. and Pautasso, G. and Piovesan, P. and Valisa, M. and Carnevale, D. and Esposito, B. and Giacomelli, L. and Gospodarczyk, M. and {McCarthy}, P. J. and Martin, P. and Suttrop, W. and Tardocchi, M. and Teschke, M. and Team, the {ASDEX} Upgrade and Team, the {EUROfusion} {MST}1},
	urldate = {2023-05-08},
	date = {2017-11},
	langid = {english},
	note = {Publisher: {IOP} Publishing},
}

@article{zeng_experimental_2013,
	title = {Experimental Observation of a Magnetic-Turbulence Threshold for Runaway-Electron Generation in the {TEXTOR} Tokamak},
	volume = {110},
	url = {https://link.aps.org/doi/10.1103/PhysRevLett.110.235003},
	doi = {10.1103/PhysRevLett.110.235003},
	pages = {235003},
	number = {23},
	journaltitle = {Physical Review Letters},
	shortjournal = {Phys. Rev. Lett.},
	author = {Zeng, L. and Koslowski, H. R. and Liang, Y. and Lvovskiy, A. and Lehnen, M. and Nicolai, D. and Pearson, J. and Rack, M. and Jaegers, H. and Finken, K. H. and Wongrach, K. and Xu, Y. and {the TEXTOR team}},
	urldate = {2023-05-08},
	date = {2013-06-05},
	note = {Publisher: American Physical Society},
}

@article{sommariva_electron_2018,
	title = {Electron acceleration in a {JET} disruption simulation},
	volume = {58},
	issn = {0029-5515},
	url = {https://dx.doi.org/10.1088/1741-4326/aad47d},
	doi = {10.1088/1741-4326/aad47d},
	pages = {106022},
	number = {10},
	journaltitle = {Nuclear Fusion},
	shortjournal = {Nucl. Fusion},
	author = {Sommariva, C. and Nardon, E. and Beyer, P. and Hoelzl, M. and Huijsmans, G. T. A. and Contributors, J. E. T.},
	urldate = {2023-05-08},
	date = {2018-08},
	langid = {english},
	note = {Publisher: {IOP} Publishing},
}

@article{svenningsson_hot-tail_2021,
	title = {Hot-Tail Runaway Seed Landscape during the Thermal Quench in Tokamaks},
	volume = {127},
	url = {https://link.aps.org/doi/10.1103/PhysRevLett.127.035001},
	doi = {10.1103/PhysRevLett.127.035001},
	pages = {035001},
	number = {3},
	journaltitle = {Physical Review Letters},
	shortjournal = {Phys. Rev. Lett.},
	author = {Svenningsson, Ida and Embreus, Ola and Hoppe, Mathias and Newton, Sarah L. and Fülöp, Tünde},
	urldate = {2023-05-08},
	date = {2021-07-14},
	note = {Publisher: American Physical Society},
}

@article{aleynikov_generation_2017,
	title = {Generation of runaway electrons during the thermal quench in tokamaks},
	volume = {57},
	issn = {0029-5515},
	url = {https://dx.doi.org/10.1088/1741-4326/aa5895},
	doi = {10.1088/1741-4326/aa5895},
	pages = {046009},
	number = {4},
	journaltitle = {Nuclear Fusion},
	shortjournal = {Nucl. Fusion},
	author = {Aleynikov, Pavel and Breizman, Boris N.},
	urldate = {2023-05-08},
	date = {2017-02},
	langid = {english},
	note = {Publisher: {IOP} Publishing},
}

@article{hollmann_study_2017,
	title = {Study of Z scaling of runaway electron plateau final loss energy deposition into wall of {DIII}-D},
	volume = {24},
	issn = {1070-664X},
	url = {https://doi.org/10.1063/1.4985086},
	doi = {10.1063/1.4985086},
	pages = {062505},
	number = {6},
	journaltitle = {Physics of Plasmas},
	shortjournal = {Physics of Plasmas},
	author = {Hollmann, E. M. and Commaux, N. and Eidietis, N. W. and Lasnier, C. J. and Rudakov, D. L. and Shiraki, D. and Cooper, C. and Martin-Solis, J. R. and Parks, P. B. and Paz-Soldan, C.},
	urldate = {2023-05-08},
	date = {2017-06-12},
}

@article{izzo_analysis_2012,
	title = {Analysis of shot-to-shot variability in post-disruption runaway electron currents for diverted {DIII}-D discharges},
	volume = {54},
	issn = {0741-3335},
	url = {https://dx.doi.org/10.1088/0741-3335/54/9/095002},
	doi = {10.1088/0741-3335/54/9/095002},
	pages = {095002},
	number = {9},
	journaltitle = {Plasma Physics and Controlled Fusion},
	shortjournal = {Plasma Phys. Control. Fusion},
	author = {Izzo, V. A. and Humphreys, D. A. and Kornbluth, M.},
	urldate = {2023-05-08},
	date = {2012-07},
	langid = {english},
	note = {Publisher: {IOP} Publishing},
}

@article{sweeney_mhd_2020,
	title = {{MHD} stability and disruptions in the {SPARC} tokamak},
	volume = {86},
	issn = {0022-3778, 1469-7807},
	url = {https://www.cambridge.org/core/journals/journal-of-plasma-physics/article/mhd-stability-and-disruptions-in-the-sparc-tokamak/908C6788C0D625C5DDF335DBD9A17476},
	doi = {10.1017/S0022377820001129},
	pages = {865860507},
	number = {5},
	journaltitle = {Journal of Plasma Physics},
	author = {Sweeney, R. and Creely, A. J. and Doody, J. and Fülöp, T. and Garnier, D. T. and Granetz, R. and Greenwald, M. and Hesslow, L. and Irby, J. and Izzo, V. A. and Haye, R. J. La and Logan, N. C. and Montes, K. and Paz-Soldan, C. and Rea, C. and Tinguely, R. A. and Vallhagen, O. and Zhu, J.},
	urldate = {2023-05-08},
	date = {2020-10},
	langid = {english},
	note = {Publisher: Cambridge University Press},
}

@article{tinguely_modeling_2021,
	title = {Modeling the complete prevention of disruption-generated runaway electron beam formation with a passive 3D coil in {SPARC}},
	volume = {61},
	issn = {0029-5515},
	url = {https://dx.doi.org/10.1088/1741-4326/ac31d7},
	doi = {10.1088/1741-4326/ac31d7},
	pages = {124003},
	number = {12},
	journaltitle = {Nuclear Fusion},
	shortjournal = {Nucl. Fusion},
	author = {Tinguely, R. A. and Izzo, V. A. and Garnier, D. T. and Sundström, A. and Särkimäki, K. and Embréus, O. and Fülöp, T. and Granetz, R. S. and Hoppe, M. and Pusztai, I. and Sweeney, R.},
	urldate = {2023-05-08},
	date = {2021-11},
	langid = {english},
	note = {Publisher: {IOP} Publishing},
}

@article{rodriguez-fernandez_overview_2022,
	title = {Overview of the {SPARC} physics basis towards the exploration of burning-plasma regimes in high-field, compact tokamaks},
	volume = {62},
	issn = {0029-5515},
	url = {https://dx.doi.org/10.1088/1741-4326/ac1654},
	doi = {10.1088/1741-4326/ac1654},
	pages = {042003},
	number = {4},
	journaltitle = {Nuclear Fusion},
	shortjournal = {Nucl. Fusion},
	author = {Rodriguez-Fernandez, P. and Creely, A. J. and Greenwald, M. J. and Brunner, D. and Ballinger, S. B. and Chrobak, C. P. and Garnier, D. T. and Granetz, R. and Hartwig, Z. S. and Howard, N. T. and Hughes, J. W. and Irby, J. H. and Izzo, V. A. and Kuang, A. Q. and Lin, Y. and Marmar, E. S. and Mumgaard, R. T. and Rea, C. and Reinke, M. L. and Riccardo, V. and Rice, J. E. and Scott, S. D. and Sorbom, B. N. and Stillerman, J. A. and Sweeney, R. and Tinguely, R. A. and Whyte, D. G. and Wright, J. C. and Yuryev, D. V.},
	urldate = {2023-05-08},
	date = {2022-03},
	langid = {english},
	note = {Publisher: {IOP} Publishing},
}

@article{izzo_runaway_2022,
	title = {Runaway electron deconfinement in {SPARC} and {DIII}-D by a passive 3D coil},
	volume = {62},
	issn = {0029-5515},
	url = {https://dx.doi.org/10.1088/1741-4326/ac83d8},
	doi = {10.1088/1741-4326/ac83d8},
	pages = {096029},
	number = {9},
	journaltitle = {Nuclear Fusion},
	shortjournal = {Nucl. Fusion},
	author = {Izzo, V. A. and Pusztai, I. and Särkimäki, K. and Sundström, A. and Garnier, D. T. and Weisberg, D. and Tinguely, R. A. and Paz-Soldan, C. and Granetz, R. S. and Sweeney, R.},
	urldate = {2023-05-08},
	date = {2022-08},
	langid = {english},
	note = {Publisher: {IOP} Publishing},
}

@article{tinguely_minimum_2023,
	title = {On the minimum transport required to passively suppress runaway electrons in {SPARC} disruptions},
	volume = {65},
	issn = {0741-3335},
	url = {https://dx.doi.org/10.1088/1361-6587/acb083},
	doi = {10.1088/1361-6587/acb083},
	pages = {034002},
	number = {3},
	journaltitle = {Plasma Physics and Controlled Fusion},
	shortjournal = {Plasma Phys. Control. Fusion},
	author = {Tinguely, R. A. and Pusztai, I. and Izzo, V. A. and Särkimäki, K. and Fülöp, T. and Garnier, D. T. and Granetz, R. S. and Hoppe, M. and Paz-Soldan, C. and Sundström, A. and Sweeney, R.},
	urldate = {2023-05-08},
	date = {2023-01},
	langid = {english},
	note = {Publisher: {IOP} Publishing},
}

@article{greenwald_status_2020,
	title = {Status of the {SPARC} physics basis},
	volume = {86},
	issn = {0022-3778, 1469-7807},
	url = {https://www.cambridge.org/core/journals/journal-of-plasma-physics/article/status-of-the-sparc-physics-basis/B21625B93C0654B955B776566C96DF6B},
	doi = {10.1017/S0022377820001063},
	pages = {861860501},
	number = {5},
	journaltitle = {Journal of Plasma Physics},
	author = {Greenwald, Martin},
	urldate = {2023-05-08},
	date = {2020-10},
	langid = {english},
	note = {Publisher: Cambridge University Press},
}

@article{chen_suppression_2018,
	title = {Suppression of runaway electrons by mode locking during disruptions on J-{TEXT}},
	volume = {58},
	issn = {0029-5515},
	url = {https://dx.doi.org/10.1088/1741-4326/aab2fc},
	doi = {10.1088/1741-4326/aab2fc},
	pages = {082002},
	number = {8},
	journaltitle = {Nuclear Fusion},
	shortjournal = {Nucl. Fusion},
	author = {Chen, Z. Y. and Lin, Z. F. and Huang, D. W. and Tong, R. H. and Hu, Q. M. and Wei, Y. N. and Yan, W. and Dai, A. J. and Zhang, X. Q. and Rao, B. and Yang, Z. J. and Gao, L. and Dong, Y. B. and Zeng, L. and Ding, Y. H. and Wang, Z. J. and Zhang, M. and Zhuang, G. and Liang, Y. and Pan, Y. and Jiang, Z. H. and Team, J.-{TEXT}},
	urldate = {2023-05-08},
	date = {2018-06},
	langid = {english},
	note = {Publisher: {IOP} Publishing},
}

@article{commaux_novel_2011,
	title = {Novel rapid shutdown strategies for runaway electron suppression in {DIII}-D},
	volume = {51},
	issn = {0029-5515},
	url = {https://dx.doi.org/10.1088/0029-5515/51/10/103001},
	doi = {10.1088/0029-5515/51/10/103001},
	pages = {103001},
	number = {10},
	journaltitle = {Nuclear Fusion},
	shortjournal = {Nucl. Fusion},
	author = {Commaux, N. and Baylor, L. R. and Combs, S. K. and Eidietis, N. W. and Evans, T. E. and Foust, C. R. and Hollmann, E. M. and Humphreys, D. A. and Izzo, V. A. and James, A. N. and Jernigan, T. C. and Meitner, S. J. and Parks, P. B. and Wesley, J. C. and Yu, J. H.},
	urldate = {2023-05-08},
	date = {2011-08},
	langid = {english},
}

@article{hollmann_experiments_2010,
	title = {Experiments in {DIII}-D toward achieving rapid shutdown with runaway electron suppressiona)},
	volume = {17},
	issn = {1070-664X},
	url = {https://doi.org/10.1063/1.3309426},
	doi = {10.1063/1.3309426},
	pages = {056117},
	number = {5},
	journaltitle = {Physics of Plasmas},
	shortjournal = {Physics of Plasmas},
	author = {Hollmann, E. M. and Commaux, N. and Eidietis, N. W. and Evans, T. E. and Humphreys, D. A. and James, A. N. and Jernigan, T. C. and Parks, P. B. and Strait, E. J. and Wesley, J. C. and Yu, J. H. and Austin, M. E. and Baylor, L. R. and Brooks, N. H. and Izzo, V. A. and Jackson, G. L. and van Zeeland, M. A. and Wu, W.},
	urldate = {2023-05-08},
	date = {2010-05-06},
}

@article{eidietis_control_2012,
	title = {Control of post-disruption runaway electron beams in {DIII}-Da)},
	volume = {19},
	issn = {1070-664X},
	url = {https://doi.org/10.1063/1.3695000},
	doi = {10.1063/1.3695000},
	pages = {056109},
	number = {5},
	journaltitle = {Physics of Plasmas},
	shortjournal = {Physics of Plasmas},
	author = {Eidietis, N. W. and Commaux, N. and Hollmann, E. M. and Humphreys, D. A. and Jernigan, T. C. and Moyer, R. A. and Strait, E. J. and {VanZeeland}, M. A. and Wesley, J. C. and Yu, J. H.},
	urldate = {2023-05-08},
	date = {2012-03-28},
}

@article{taylor_disruption_1999,
	title = {Disruption mitigation studies in {DIII}-D},
	volume = {6},
	issn = {1070-664X},
	url = {https://doi.org/10.1063/1.873445},
	doi = {10.1063/1.873445},
	pages = {1872--1879},
	number = {5},
	journaltitle = {Physics of Plasmas},
	shortjournal = {Physics of Plasmas},
	author = {Taylor, P. L. and Kellman, A. G. and Evans, T. E. and Gray, D. S. and Humphreys, D. A. and Hyatt, A. W. and Jernigan, T. C. and Lee, R. L. and Leuer, J. A. and Luckhardt, S. C. and Parks, P. B. and Schaffer, M. J. and Whyte, D. G. and Zhang, J.},
	urldate = {2023-05-08},
	date = {1999-05-01},
}

@article{shiraki_dissipation_2018,
	title = {Dissipation of post-disruption runaway electron plateaus by shattered pellet injection in {DIII}-D},
	volume = {58},
	issn = {0029-5515},
	url = {https://dx.doi.org/10.1088/1741-4326/aab0d6},
	doi = {10.1088/1741-4326/aab0d6},
	pages = {056006},
	number = {5},
	journaltitle = {Nuclear Fusion},
	shortjournal = {Nucl. Fusion},
	author = {Shiraki, D. and Commaux, N. and Baylor, L. R. and Cooper, C. M. and Eidietis, N. W. and Hollmann, E. M. and Paz-Soldan, C. and Combs, S. K. and Meitner, S. J.},
	urldate = {2023-05-08},
	date = {2018-03},
	langid = {english},
	note = {Publisher: {IOP} Publishing},
}

@article{luxon_design_2002,
	title = {A design retrospective of the {DIII}-D tokamak},
	volume = {42},
	issn = {0029-5515},
	url = {https://dx.doi.org/10.1088/0029-5515/42/5/313},
	doi = {10.1088/0029-5515/42/5/313},
	pages = {614},
	number = {5},
	journaltitle = {Nuclear Fusion},
	shortjournal = {Nucl. Fusion},
	author = {Luxon, J. L.},
	urldate = {2023-05-08},
	date = {2002-05},
	langid = {english},
}

@article{battey_simultaneous_2023,
	title = {Simultaneous stabilization and control of the n = 1 and n = 2 resistive wall mode},
	volume = {63},
	rights = {All rights reserved},
	issn = {0029-5515},
	url = {https://dx.doi.org/10.1088/1741-4326/accd81},
	doi = {10.1088/1741-4326/accd81},
	pages = {066025},
	number = {6},
	journaltitle = {Nuclear Fusion},
	shortjournal = {Nucl. Fusion},
	author = {Battey, A. F. and Hanson, J. M. and Bialek, J. and Turco, F. and Navratil, G. A. and Logan, N. C.},
	urldate = {2023-05-08},
	date = {2023-04},
	langid = {english},
	note = {Publisher: {IOP} Publishing},
}

@article{weisberg_passive_2021,
	title = {Passive deconfinement of runaway electrons using an in-vessel helical coil},
	volume = {61},
	issn = {0029-5515},
	url = {https://dx.doi.org/10.1088/1741-4326/ac2279},
	doi = {10.1088/1741-4326/ac2279},
	pages = {106033},
	number = {10},
	journaltitle = {Nuclear Fusion},
	shortjournal = {Nucl. Fusion},
	author = {Weisberg, D. B. and Paz-Soldan, C. and Liu, Y. Q. and Welander, A. and Dunn, C.},
	urldate = {2023-05-08},
	date = {2021-09},
	langid = {english},
	note = {Publisher: {IOP} Publishing},
}

@article{boozer_two_2011,
	title = {Two beneficial non-axisymmetric perturbations to tokamaks},
	volume = {53},
	issn = {0741-3335},
	url = {https://dx.doi.org/10.1088/0741-3335/53/8/084002},
	doi = {10.1088/0741-3335/53/8/084002},
	pages = {084002},
	number = {8},
	journaltitle = {Plasma Physics and Controlled Fusion},
	shortjournal = {Plasma Phys. Control. Fusion},
	author = {Boozer, Allen H.},
	urldate = {2023-05-08},
	date = {2011-05},
	langid = {english},
}

@article{smith_passive_2013,
	title = {Passive runaway electron suppression in tokamak disruptions},
	volume = {20},
	issn = {1070-664X},
	url = {https://doi.org/10.1063/1.4813255},
	doi = {10.1063/1.4813255},
	pages = {072505},
	number = {7},
	journaltitle = {Physics of Plasmas},
	shortjournal = {Physics of Plasmas},
	author = {Smith, H. M. and Boozer, A. H. and Helander, P.},
	urldate = {2023-05-08},
	date = {2013-07-24},
}

@article{jiang_simulation_2016,
	title = {Simulation of runaway electrons, transport affected by J-{TEXT} resonant magnetic perturbation},
	volume = {56},
	issn = {0029-5515},
	url = {https://dx.doi.org/10.1088/0029-5515/56/9/092012},
	doi = {10.1088/0029-5515/56/9/092012},
	pages = {092012},
	number = {9},
	journaltitle = {Nuclear Fusion},
	shortjournal = {Nucl. Fusion},
	author = {Jiang, Z. H. and Wang, X. H. and Chen, Z. Y. and Huang, D. W. and Sun, X. F. and Xu, T. and Zhuang, G.},
	urldate = {2023-05-08},
	date = {2016-07},
	langid = {english},
	note = {Publisher: {IOP} Publishing},
}

@article{mlynar_runaway_2018,
	title = {Runaway electron experiments at {COMPASS} in support of the {EUROfusion} {ITER} physics research},
	volume = {61},
	issn = {0741-3335},
	url = {https://dx.doi.org/10.1088/1361-6587/aae04a},
	doi = {10.1088/1361-6587/aae04a},
	pages = {014010},
	number = {1},
	journaltitle = {Plasma Physics and Controlled Fusion},
	shortjournal = {Plasma Phys. Control. Fusion},
	author = {Mlynar, J. and Ficker, O. and Macusova, E. and Markovic, T. and Naydenkova, D. and Papp, G. and Urban, J. and Vlainic, M. and Vondracek, P. and Weinzettl, V. and Bogar, O. and Bren, D. and Carnevale, D. and Casolari, A. and Cerovsky, J. and Farnik, M. and Gobbin, M. and Gospodarczyk, M. and Hron, M. and Kulhanek, P. and Havlicek, J. and Havranek, A. and Imrisek, M. and Jakubowski, M. and Lamas, N. and Linhart, V. and Malinowski, K. and Marcisovsky, M. and Matveeva, E. and Panek, R. and Plyusnin, V. V. and Rabinski, M. and Svoboda, V. and Svihra, P. and Varju, J. and Zebrowski, J. and Team, the {COMPASS} and Team, the {EUROfusion} {MST}1},
	urldate = {2023-05-08},
	date = {2018-11},
	langid = {english},
	note = {Publisher: {IOP} Publishing},
}

@article{papp_runaway_2011,
	title = {Runaway electron drift orbits in magnetostatic perturbed fields},
	volume = {51},
	issn = {0029-5515},
	url = {https://dx.doi.org/10.1088/0029-5515/51/4/043004},
	doi = {10.1088/0029-5515/51/4/043004},
	pages = {043004},
	number = {4},
	journaltitle = {Nuclear Fusion},
	shortjournal = {Nucl. Fusion},
	author = {Papp, G. and Drevlak, M. and Fülöp, T. and Helander, P.},
	urldate = {2023-05-08},
	date = {2011-03},
	langid = {english},
}

@article{xue_hot_2019,
	title = {Hot {VDE} investigation of the negative triangularity plasmas based on {HL}-2M tokamak},
	volume = {143},
	issn = {0920-3796},
	url = {https://www.sciencedirect.com/science/article/pii/S0920379619304363},
	doi = {10.1016/j.fusengdes.2019.03.103},
	pages = {48--58},
	journaltitle = {Fusion Engineering and Design},
	shortjournal = {Fusion Engineering and Design},
	author = {Xue, L. and Zheng, G. Y. and Duan, X. R. and Liu, Y. Q. and Hoang, G. T and Li, J. X. and Dokuka, V. N. and Lukash, V. E. and Khayrutdinov, R. R.},
	urldate = {2023-05-09},
	date = {2019-06-01},
	langid = {english},
}

@article{nakamura_mechanism_1996,
	title = {Mechanism of vertical displacement events in {JT}-60U disruptive discharges},
	volume = {36},
	issn = {0029-5515},
	doi = {10.1088/0029-5515/36/5/I10},
	pages = {643--656},
	number = {5},
	journaltitle = {Nuclear Fusion},
	author = {Nakamura, Y. and Yoshino, R. and Neyatani, Y. and Tsunematsu, T. and Azumi, M. and Pomphrey, N. and Jardin, S.C.},
	date = {1996},
}

@article{nakamura_p-collapse-induced_1996,
	title = {βp-collapse-induced vertical displacement event in high βp tokamak disruption},
	volume = {38},
	issn = {0741-3335},
	doi = {10.1088/0741-3335/38/10/007},
	pages = {1791--1804},
	number = {10},
	journaltitle = {Plasma Physics and Controlled Fusion},
	author = {Nakamura, Y. and Yoshino, R. and Pomphrey, N. and Jardin, S.C.},
	date = {1996},
}

@article{fitzpatrick_simple_2009,
	title = {A simple ideal magnetohydrodynamical model of vertical disruption events in tokamaks},
	volume = {16},
	issn = {1070-664X},
	url = {https://doi.org/10.1063/1.3068467},
	doi = {10.1063/1.3068467},
	pages = {012506},
	number = {1},
	journaltitle = {Physics of Plasmas},
	shortjournal = {Physics of Plasmas},
	author = {Fitzpatrick, R.},
	urldate = {2023-05-09},
	date = {2009-01-27},
}

@article{gruber_vertical_1993,
	title = {Vertical displacement events and halo currents},
	volume = {35},
	issn = {0741-3335},
	doi = {10.1088/0741-3335/35/SB/015},
	pages = {B191--B204},
	issue = {{SB}},
	journaltitle = {Plasma Physics and Controlled Fusion},
	author = {Gruber, O. and Lackner, K. and Pautasso, G. and Seidel, U. and Streibl, B.},
	date = {1993},
}

@article{lukash_influence_2007,
	title = {Influence of plasma opacity on current decay after disruptions in tokamaks},
	volume = {47},
	issn = {0029-5515},
	url = {https://dx.doi.org/10.1088/0029-5515/47/11/009},
	doi = {10.1088/0029-5515/47/11/009},
	pages = {1476},
	number = {11},
	journaltitle = {Nuclear Fusion},
	shortjournal = {Nucl. Fusion},
	author = {Lukash, V. E. and Mineev, A. B. and Morozov, D. Kh},
	urldate = {2023-05-09},
	date = {2007-10},
	langid = {english},
}

@article{martin-solis_formation_2017,
	title = {Formation and termination of runaway beams in {ITER} disruptions},
	volume = {57},
	issn = {0029-5515},
	url = {https://dx.doi.org/10.1088/1741-4326/aa6939},
	doi = {10.1088/1741-4326/aa6939},
	pages = {066025},
	number = {6},
	journaltitle = {Nuclear Fusion},
	shortjournal = {Nucl. Fusion},
	author = {Martín-Solís, J. R. and Loarte, A. and Lehnen, M.},
	urldate = {2023-05-09},
	date = {2017-04},
	langid = {english},
	note = {Publisher: {IOP} Publishing},
}

@article{boozer_runaway_2017,
	title = {Runaway electrons and {ITER}},
	volume = {57},
	issn = {0029-5515},
	url = {https://dx.doi.org/10.1088/1741-4326/aa6355},
	doi = {10.1088/1741-4326/aa6355},
	pages = {056018},
	number = {5},
	journaltitle = {Nuclear Fusion},
	shortjournal = {Nucl. Fusion},
	author = {Boozer, Allen H.},
	urldate = {2023-05-09},
	date = {2017-03},
	langid = {english},
	note = {Publisher: {IOP} Publishing},
}

@article{hesslow_influence_2019,
	title = {Influence of massive material injection on avalanche runaway generation during tokamak disruptions},
	volume = {59},
	issn = {0029-5515},
	url = {https://dx.doi.org/10.1088/1741-4326/ab26c2},
	doi = {10.1088/1741-4326/ab26c2},
	pages = {084004},
	number = {8},
	journaltitle = {Nuclear Fusion},
	shortjournal = {Nucl. Fusion},
	author = {Hesslow, L. and Embréus, O. and Vallhagen, O. and Fülöp, T.},
	urldate = {2023-05-09},
	date = {2019-06},
	langid = {english},
	note = {Publisher: {IOP} Publishing},
}

@article{nilsson_kinetic_2015,
	title = {Kinetic modelling of runaway electron avalanches in tokamak plasmas},
	volume = {57},
	issn = {0741-3335},
	url = {https://dx.doi.org/10.1088/0741-3335/57/9/095006},
	doi = {10.1088/0741-3335/57/9/095006},
	pages = {095006},
	number = {9},
	journaltitle = {Plasma Physics and Controlled Fusion},
	shortjournal = {Plasma Phys. Control. Fusion},
	author = {Nilsson, E. and Decker, J. and Peysson, Y. and Granetz, R. S. and Saint-Laurent, F. and Vlainic, M.},
	urldate = {2023-05-09},
	date = {2015-07},
	langid = {english},
	note = {Publisher: {IOP} Publishing},
}

@article{mcdevitt_avalanche_2019,
	title = {Avalanche mechanism for runaway electron amplification in a tokamak plasma},
	volume = {61},
	issn = {0741-3335},
	url = {https://dx.doi.org/10.1088/1361-6587/ab0d6d},
	doi = {10.1088/1361-6587/ab0d6d},
	pages = {054008},
	number = {5},
	journaltitle = {Plasma Physics and Controlled Fusion},
	shortjournal = {Plasma Phys. Control. Fusion},
	author = {{McDevitt}, Christopher J. and Guo, Zehua and Tang, Xian-Zhu},
	urldate = {2023-05-09},
	date = {2019-04},
	langid = {english},
	note = {Publisher: {IOP} Publishing},
}

@article{saint-laurent_overview_2013,
	title = {Overview of Runaway Electron Control and Mitigation Experiments on Tore Supra and Lessons Learned in View of {ITER}},
	volume = {64},
	issn = {1536-1055},
	url = {https://doi.org/10.13182/FST13-A24090},
	doi = {10.13182/FST13-A24090},
	pages = {711--718},
	number = {4},
	journaltitle = {Fusion Science and Technology},
	author = {Saint-Laurent, F. and Martin, G. and Alarcon, T. and Le Luyer, A. and Parks, P. B. and Pastor, P. and Putvinski, S. and Reux, C. and Bucalossi, J. and Bremond, S. and Moreau, P.H.},
	urldate = {2023-05-09},
	date = {2013-11-01},
	note = {Publisher: Taylor \& Francis
\_eprint: https://doi.org/10.13182/{FST}13-A24090},
}

@article{iter_physics_expert_group_on_disruptions_chapter_1999,
	title = {Chapter 3: {MHD} stability, operational limits and disruptions},
	volume = {39},
	issn = {0029-5515},
	url = {https://dx.doi.org/10.1088/0029-5515/39/12/303},
	doi = {10.1088/0029-5515/39/12/303},
	shorttitle = {Chapter 3},
	pages = {2251},
	number = {12},
	journaltitle = {Nuclear Fusion},
	shortjournal = {Nucl. Fusion},
	author = {ITER Physics Expert Group on Disruptions, Plasma Control,  and {MHD} and Editors, ITER Physics Basis},
	urldate = {2023-05-09},
	date = {1999-12},
	langid = {english},
}

@article{hansen_numerical_2015,
	title = {Numerical studies and metric development for validation of magnetohydrodynamic models on the {HIT}-{SI} experimenta)},
	volume = {22},
	issn = {1070-664X},
	url = {https://doi.org/10.1063/1.4919277},
	doi = {10.1063/1.4919277},
	pages = {056105},
	number = {5},
	journaltitle = {Physics of Plasmas},
	shortjournal = {Physics of Plasmas},
	author = {Hansen, C. and Victor, B. and Morgan, K. and Jarboe, T. and Hossack, A. and Marklin, G. and Nelson, B. A. and Sutherland, D.},
	urldate = {2023-05-09},
	date = {2015-04-27},
}

@thesis{hansen_mhd_2014,
	title = {{MHD} Modeling in Complex 3D Geometries: Towards Predictive Simulation of {SIHI} Current Drive},
	rights = {Copyright is held by the individual authors.},
	url = {https://digital.lib.washington.edu:443/researchworks/handle/1773/25420},
	shorttitle = {{MHD} Modeling in Complex 3D Geometries},
	type = {Thesis},
	author = {Hansen, Christopher James},
	urldate = {2023-05-09},
	date = {2014-04-30},
	langid = {american},
	note = {Accepted: 2014-04-30T16:21:50Z},
}

@online{noauthor_hdf5_nodate,
	title = {{HDF}5: {HDF}5 Reference Manual},
	url = {https://docs.hdfgroup.org/hdf5/develop/_r_m.html},
	urldate = {2023-05-09},
}

@online{noauthor_xdmf_nodate,
	title = {{XDMF} Model and Format - {XdmfWeb}},
	url = {https://www.xdmf.org/index.php/XDMF_Model_and_Format},
	urldate = {2023-05-09},
}

@online{noauthor_citing_nodate,
	title = {Citing {VisIt}},
	url = {https://visit-dav.github.io/visit-website/citing-visit/},
	titleaddon = {{VisIt} Home},
	urldate = {2023-05-09},
	langid = {english},
}

@article{maxwell_james_c_faradays_1855,
	title = {On Faraday's Lines of Force},
	pages = {27--83},
	journaltitle = {Cambridge Philosophical Society Transactions},
	author = {{Maxwell, James C.}},
	date = {1855},
}

@article{wesson_disruptions_1989,
	title = {Disruptions in {JET}},
	volume = {29},
	issn = {0029-5515},
	url = {https://dx.doi.org/10.1088/0029-5515/29/4/009},
	doi = {10.1088/0029-5515/29/4/009},
	pages = {641},
	number = {4},
	journaltitle = {Nuclear Fusion},
	shortjournal = {Nucl. Fusion},
	author = {Wesson, J. A. and Gill, R. D. and Hugon, M. and Schüller, F. C. and Snipes, J. A. and Ward, D. J. and Bartlett, D. V. and Campbell, D. J. and Duperrex, P. A. and Edwards, A. W. and Granetz, R. S. and Gottardi, N. A. O. and Hender, T. C. and Lazzaro, E. and Lomas, P. J. and Cardozo, N. Lopes and Mast, K. F. and Nave, M. F. F. and Salmon, N. A. and Smeulders, P. and Thomas, P. R. and Tubbing, B. J. D. and Turner, M. F. and Weller, A.},
	urldate = {2023-05-09},
	date = {1989-04},
	langid = {english},
}

@article{pustovitov_models_2022,
	title = {Models and scalings for the disruption forces in tokamaks},
	volume = {62},
	issn = {0029-5515},
	url = {https://dx.doi.org/10.1088/1741-4326/ac3fe9},
	doi = {10.1088/1741-4326/ac3fe9},
	pages = {026036},
	number = {2},
	journaltitle = {Nuclear Fusion},
	shortjournal = {Nucl. Fusion},
	author = {Pustovitov, V. D.},
	urldate = {2023-05-09},
	date = {2022-01},
	langid = {english},
	note = {Publisher: {IOP} Publishing},
}

@article{boozer_equations_1998,
	title = {Equations for studies of feedback stabilization},
	volume = {5},
	issn = {1070-664X},
	url = {https://doi.org/10.1063/1.873048},
	doi = {10.1063/1.873048},
	pages = {3350--3357},
	number = {9},
	journaltitle = {Physics of Plasmas},
	shortjournal = {Physics of Plasmas},
	author = {Boozer, Allen H.},
	urldate = {2023-05-09},
	date = {1998-09-01},
}

@article{artola_3d_2021,
	title = {3D simulations of vertical displacement events in tokamaks: A benchmark of M3D-C1, {NIMROD}, and {JOREK}},
	volume = {28},
	issn = {1070-664X},
	url = {https://doi.org/10.1063/5.0037115},
	doi = {10.1063/5.0037115},
	shorttitle = {3D simulations of vertical displacement events in tokamaks},
	pages = {052511},
	number = {5},
	journaltitle = {Physics of Plasmas},
	shortjournal = {Physics of Plasmas},
	author = {Artola, F. J. and Sovinec, C. R. and Jardin, S. C. and Hoelzl, M. and Krebs, I. and Clauser, C.},
	urldate = {2023-05-09},
	date = {2021-05-24},
}

@article{jackson_iter_2008,
	title = {{ITER} startup studies in the {DIII}-D tokamak},
	volume = {48},
	issn = {0029-5515},
	url = {https://dx.doi.org/10.1088/0029-5515/48/12/125002},
	doi = {10.1088/0029-5515/48/12/125002},
	pages = {125002},
	number = {12},
	journaltitle = {Nuclear Fusion},
	shortjournal = {Nucl. Fusion},
	author = {Jackson, G. L. and Casper, T. A. and Luce, T. C. and Humphreys, D. A. and Ferron, J. R. and Hyatt, A. W. and Lazarus, E. A. and Moyer, R. A. and Petrie, T. W. and Rudakov, D. L. and West, W. P.},
	urldate = {2023-05-10},
	date = {2008-11},
	langid = {english},
}

@article{jackson_simulating_2009,
	title = {Simulating {ITER} plasma startup and rampdown scenarios in the {DIII}-D tokamak},
	volume = {49},
	issn = {0029-5515},
	url = {https://dx.doi.org/10.1088/0029-5515/49/11/115027},
	doi = {10.1088/0029-5515/49/11/115027},
	pages = {115027},
	number = {11},
	journaltitle = {Nuclear Fusion},
	shortjournal = {Nucl. Fusion},
	author = {Jackson, G. L. and Casper, T. A. and Luce, T. C. and Humphreys, D. A. and Ferron, J. R. and Hyatt, A. W. and Leuer, J. A. and Petrie, T. W. and Turco, F. and West, W. P.},
	urldate = {2023-05-10},
	date = {2009-10},
	langid = {english},
}

@article{ferguson_complete_1994,
	title = {A complete linear discretization for calculating the magnetic field using the boundary element method},
	volume = {41},
	doi = {10.1109/10.293220},
	pages = {455--460},
	number = {5},
	journaltitle = {{IEEE} Transactions on Biomedical Engineering},
	author = {Ferguson, A.S. and Zhang, Xu and Stroink, G.},
	date = {1994},
}

@article{humphreys_analytic_1999,
	title = {Analytic modeling of axisymmetric disruption halo currents},
	volume = {6},
	issn = {1070-664X},
	url = {https://doi.org/10.1063/1.873231},
	doi = {10.1063/1.873231},
	pages = {2742--2756},
	number = {7},
	journaltitle = {Physics of Plasmas},
	shortjournal = {Physics of Plasmas},
	author = {Humphreys, D. A. and Kellman, A. G.},
	urldate = {2023-05-20},
	date = {1999-07-01},
}

@article{eidietis_diffusive_2011,
	title = {A diffusive model for halo width growth during vertical displacement events},
	volume = {51},
	issn = {0029-5515},
	url = {https://dx.doi.org/10.1088/0029-5515/51/7/073034},
	doi = {10.1088/0029-5515/51/7/073034},
	pages = {073034},
	number = {7},
	journaltitle = {Nuclear Fusion},
	shortjournal = {Nucl. Fusion},
	author = {Eidietis, N. W. and Humphreys, D. A.},
	urldate = {2023-05-20},
	date = {2011-06},
	langid = {english},
}

@misc{merkel_linear_2015,
	title = {Linear {MHD} stability studies with the {STARWALL} code},
	url = {http://arxiv.org/abs/1508.04911},
	doi = {10.48550/arXiv.1508.04911},
	number = {{arXiv}:1508.04911},
	publisher = {{arXiv}},
	author = {Merkel, P. and Strumberger, E.},
	urldate = {2023-05-30},
	date = {2015-08-20},
	eprinttype = {arxiv},
	eprint = {1508.04911 [physics]},
}
\end{document}